\documentclass[a4paper,12pt]{article}
\usepackage{jcappub} % for details on the use of the package, please
\usepackage{cite}
\usepackage{graphicx}
\usepackage{enumerate}
\usepackage{cancel}
\usepackage{color}
\usepackage{graphicx}
\usepackage{amsmath}
\usepackage{amsfonts}
\usepackage{amssymb}
\usepackage{rotating}
\usepackage{booktabs}
\usepackage{xcolor}
\usepackage{soul}

\def\blue#1{\textcolor{blue}{#1}}
\def\green#1{\textcolor{green}{#1}}

\def\comment#1{}

\title{\boldmath 
Higgs boson origin from a gauge symmetric theory of massive composite particles and massless $W^\pm$ and $Z^0$ bosons at the TeV scale}

\author{She-Sheng Xue}

\affiliation{ICRANet Piazzale della Repubblica, 10 -65122, Pescara, Italy
\\ Physics Department, University of Rome La Sapienza, 
Rome, Italy\\INFN, Sezione di Perugia, %Via A. Pascoli, I-06123, 
Perugia, Italy
\\ICTP-AP, University of Chinese Academy of Sciences, Beijing, China
}

\emailAdd{xue@icra.it and shesheng.xue@gmail.com} 

\abstract{
The ultraviolet completion is the Standard Model (SM) gauge-symmetric four-fermion couplings at 
the high-energy cutoff. Composite particles appear in the gauge symmetric phase in contrast with SM particles 
in the spontaneous symmetry-breaking phase. 
The critical point between the two phases is a weak first-order transition. It relates to an ultraviolet fixed point for an SM gauge symmetric theory of composite particles in the strong coupling regime. The low-energy SM realizes at an infrared fixed point in the weak coupling regime. Composite bosons dissolve into SM particles at the phase transition, and in the top-quark channel, they become a composite SM Higgs boson and three Goldstone bosons. Extrapolation of SM renormalization-group solutions to high energies implies that the 
gauge-symmetric theory of composite particles has a characteristic scale of about $5.1$ TeV. We discuss the phenomenological implications of composite SM Higgs boson in the gauge symmetry-breaking phase and massive composite bosons coupling to massless $W^\pm$ and $Z^0$ gauge bosons in the gauge symmetric phase.
}

\begin{document}
\maketitle
\flushbottom

%%%%%%%%%%%%%%%%%%%%%%%%%%%%%%%%%%%%%%%%%%%%%%%%%%%%%%%%%%%%%%%%%%%%%%%%%%%%%%%%%%%%%%%%%%%%%%%%%%%%%%%%%%%%%%%%%%%%%%%%%%%%
%%%%%%%%%%%%%%%%%%%%%%%%%%%%%%%%%%%%%%%%%%%%%%%%%%%%%%%%%%%%%%%%%%%%%%%%%%%%%%%%%%%%%%%%%%%%%%%%%%%%%%%%%%%%%%%%%%%%%%%%%%%%
%%%%%%%%%%%%%%%%%%%%%%%%%%%%%%%%%%%%%%%%%%%%%%%%%%%%%%%%%%%%%%%%%%%%%%%%%%%%%%%%%%%%%%%%%%%%%%%%%%%%%%%%%%%%%%%%%%%%%%%%%%%%
%

\section{Theoretical ultraviolet completion}\label{uv}
%\hskip0.1cm

The parity-violating gauge symmetries and spontaneous/explicit breaking of these symmetries for the $W$ and $Z$ gauge boson 
masses and the hierarchy pattern of fermion masses have been at the 
centre of conceptual elaboration. It has played a major
role in donating to Mankind the beauty of the Standard
Model (SM) and possible scenarios beyond SM for fundamental particle physics. 
A simple description is provided on the one hand by the composite Higgs-boson model
or the Nambu-Jona-Lasinio (NJL) model \cite{Nambu:1961fr} with effective four-fermion operators, and on the other by the phenomenological model of the elementary Higgs boson \cite{Englert1964, Higgs1964, Guralnik1964}. 
%The latter is not explained in the SM. 
These two models are effectively equivalent for the SM at
low energies. The ATLAS \cite{ATLAS:2012yve} 
and CMS \cite{Chatrchyan2012}
collaborations have shown the first observations of a
125 GeV scalar particle in the search for the SM Higgs
boson. 

A well-defined quantum field theory for the SM Lagrangian requires a natural regularisation at the 
ultraviolet energy cutoff $\Lambda_{\rm cut}=2\pi/a$ (the spatial lattice spacing $a$), fully preserving the SM chiral gauge symmetry.
The cutoff $\Lambda_{\rm cut}$ could be the Planck or the grand unified theory scale.
Quantum gravity or another new physics naturally provides such regularisation. 
However, the No-Go theorem \cite{Nielsen:1981xu, 1981PhLB..105..219N} shows the presence of right-handed neutrinos and 
the absence of consistent regularisation for the SM bilinear fermion Lagrangian to preserve the SM chiral gauge symmetries.
It implies SM fermions' and right-handed neutrinos' four-fermion (quadrilinear) operators at the cutoff $\Lambda_{\rm cut}$.
As a theoretical model, we adopt the four-fermion operators of the torsion-free Einstein-Cartan Lagrangian with all SM fermions and three right-handed neutrinos \cite{Xue2015, Xue:2016dpl, Xue2017}. Among four-fermion operators of Einstein-Cartan or NJL type, we consider here one for the third quark family
\begin{eqnarray}
G_{\rm cut}\Big[(\bar\psi^{ia}_Lt_{Ra})(\bar t^b_{R}\psi_{Lib}) + (\bar\psi^{ia}_Lb_{Ra})(\bar b^b_{R}\psi_{Lib})\Big],
\label{bhl}
\end{eqnarray}
where $a$ and $b$ are the colour indexes of the top and bottom quarks. The $SU_L(2)$ singlets $\psi_R^a 
= t^{a}_R, b^{a}_R$ and doublet $\psi^{ia}_L=(t^{a}_L,b^{a}_L)$ are the eigenstates of SM electroweak interactions. 
Henceforth, we suppress the colour indexes.
The effective four-fermion coupling 
$G_{\rm cut}\propto {\mathcal O}(\Lambda^{-2}_{\rm cut})$ and the dimensionless coupling $G_{\rm cut}\Lambda^{2}_{\rm cut}$ depends on the running energy scale $\mu$. Through the help of an auxiliary 
(static) scalar field $H_0$,  we rewrite the four-fermion interaction (\ref{bhl}) as 
\begin{eqnarray}
g_H^{\rm cut}(\bar \psi_L t_RH_0+ {\rm h.c.}) - M_0^2H^\dagger_0 H_0 + (t\rightarrow b).
\label{static}
\end{eqnarray}
where the bare Yukawa coupling $g_H^{\rm cut}=G_{\rm cut}\Lambda^2_{\rm cut}$. Integration over the static field $H_0$ of mass $M_0=\Lambda_{\rm cut}$ yields 
the four-fermion interaction (\ref{bhl}) up to non-dynamical constants. 

Apart from what is possible new physics at the cutoff $\Lambda_{\rm cut}$ 
explaining the origin of these effective four-fermion operators (\ref{bhl}), 
it is essential to study their interactions and ground state: (i) which dynamics these operators undergo in terms of their couplings as functions of running energy scale $\mu$; (ii) associating to these dynamics where the infrared (IR) or ultraviolet (UV) stable fix-point of physical couplings locates; (iii) in the IR or UV scaling domains (scaling regions) of these stable fixed points, 
which physically relevant operators that 
become effectively dimensional-4 ($d=4$) renormalizable
operators following renormalization-group (RG) equations (scaling laws); (iv) which ($d>4$) 
irrelevant operators, suppressed by the cutoff scale, have sizable corrections to the relevant operators. Here we mention the discussions on the triviality of the four-dimensional pure $m_\phi^2\phi^2+\lambda_\phi \phi^4$
theory, namely, whether the coupling  $\lambda_\phi\rightarrow 0$ at its fixed point for the cutoff $\Lambda_{\rm cut}\gg m_\phi$. 

Using the four-fermion operators (\ref{bhl}), Bardeen, Hill and Lindner (BHL) \cite{Bardeen1990} investigate the dynamics of spontaneous gauge symmetry breaking (SSB), and an IR fixed point $G_c$ at the weak coupling regime, where the $\beta$-function $\beta=\mu\partial G_{\rm cut}/\partial\mu>0$. In the IR scaling domain, gauge symmetries break, and the effective field theory of massive top quark, $W^\pm, Z^0$ and $\langle\bar tt\rangle$-composite Higgs is realised for the SM model at the electroweak scale $v=(\sqrt{2}G_F)^{-1/2}\approx 246$ GeV, where $G_F$ is the measured Fermi constant. However, the compositeness conditions at the scale 
$\Lambda_{\rm cut}$ do not yield the top-quark and Higgs masses consistently with experimental values.

In contrast, we investigate \cite{Xue1996, Xue2000, Xue2017} the dynamics of SM elementary fermions forming composite bosons $\Pi\propto (\bar\psi_L \psi_R)$ 
and fermions $\Psi_{L,R}\propto (\bar\psi_L \psi_R)~\psi_{L,R}$, and a UV fixed point at 
the strong coupling regime, where the $\beta$-function $\beta=\mu\partial G_{\rm cut}/\partial\mu < 0$. In the UV scaling domain, the SM chiral gauge symmetries preserve, and the effective field theory of massive composite particles and massless gauge bosons $W^\pm, Z^0$ realise at the scale $\Lambda$ ($v<\Lambda<\Lambda_{\rm cut}$). The UV fixed point relates to the critical point of the phase transition 
from the strong-coupling symmetric phase to the weak-coupling symmetry-breaking phase \cite{Xue:1999xa, Xue2014, Xue2017}.
The scenario has been generalised to the strongly-coupled Fermi liquid for the Bose-Einstein condensate \cite{Kleinert2018}.
Reference \cite {Leonardi:2018jzn} shows the basic phenomenology of composite bosons and fermions.

The weak and strong coupling regimes (\ref{bhl}) bring us into two distinct domains: the UV and IR domains. This is a non-perturbative issue. 
It is reminiscent of the QCD dynamics: asymptotic free quark states near a UV fixed point $\beta<0$ at high energies and bound hadron states near a possible IR fixed point $\beta>0$ at low energies.
In the present scenario, the main theoretical problems yet to be solved are the following. (i) The quantitative features of the strong coupling domain: SM gauge couplings, and mass spectra of composite particles. (ii) The nature of phase transition at the TeV scale: from the exact gauge symmetric phase for the massless gauge bosons and massive composite particles to the gauge-symmetry breaking phase for the SM elementary gauge bosons, Higgs boson, quarks and leptons. In this article, to address these issues,  we focus on the four composite bosons $H^\pm$ and $H^0_{t,b}$ composed by $t$ and $b$ quarks in the third quark family. We study the massive composite bosons $H$ mass terms, the effective potential $V(H^2)$ and their Yukawa coupling $g_H$ to $t$ and $b$ quarks at the TeV scale. 
We find how the gauge symmetry ground state evolves into the symmetry-breaking ground state. The results show the phase transition of spontaneous symmetry breaking is the weak first order. Upon the transition, the composite boson $H_t^0$ becomes the composite SM Higgs boson $h$, and others become the $W^\pm$ and $Z^0$ gauge bosons' longitudinal modes. We present phenomenological implications of massless gauge bosons $W^\pm$ and $Z^0$ at the TeV scale. For readers' convenience, we briefly review the results in previous publications that are necessary gradients for discussing the issue here so that the article is self-consistent and self-contain.

In Secs.~\ref{scsp} and \ref{uvcom}, we discuss the strong-coupling symmetric phase and the  
critical point of phase transition
as a UV fixed point to realise an effective theory for composite particles. In Secs.~\ref{firstorder} and \ref{weakc}, we show the weak first-order phase 
transition to the symmetry-breaking phase in the weak 
coupling regime. In Sec.~\ref{disso}, we discuss 
how composite 
particles dissolve into their constituents at the 
phase transition, leading to a composite SM Higgs boson. In Secs.~\ref{SSBS} and \ref{extrap}, we present the investigations on
the IR fixed point for the low-energy SM model 
in connection with the UV fixed point for the theory of composite particles. In the last section \ref{phenoW}, we discuss the phenomenological implications of composite SM Higgs boson in the IR domain, massive composite bosons and massless $W^\pm$ and $Z^0$ gauge bosons in the UV domain.

%\vskip0.1cm
\section
{Strong-coupling gauge symmetric phase}\label{scsp}
\hskip0.1cm 
In the strong-coupling limit $G_{\rm cut}a^{-2}\gg 1$, the theory (\ref{bhl}) is in the strong-coupling symmetric phase, 
composite particles' spectra preserve the SM gauge symmetries $SU_c(3)\times SU_L(2)\times U_Y(1)$ \cite{Xue1996, Xue:1999xa}. We calculate the propagators of composite fermion and boson fields by the strong 
coupling expansion in powers of kinetic term $1/g^{1/2}$ 
\begin{eqnarray}
S_{\rm kinetic}&=&\frac{1}{ 2ag^{1/2}}\sum_{x,\mu}\bar\psi(x)
\gamma_\mu \partial^\mu\psi(x),\quad %\partial^\mu\equiv \delta_{x,x+a_\mu}-\delta_{x,x-a_\mu}\label{rfa}
\nonumber\\
S_{\rm int}&=&\sum_{x}\left[(\bar\psi^{ia}_Lt_{Ra})(\bar t^b_{R}\psi_{Lib}) + (\bar\psi^{ia}_Lb_{Ra})(\bar b^b_{R}\psi_{Lib})\right],\label{rs2}
\end{eqnarray}
where $g\equiv G_{\rm cut}/a^4\,$ ($ga^2\gg 1$)
and scaling $
\psi(x)\,\rightarrow \psi(x)=a^2 g^{1/4}\psi(x)$.
In the lowest non-trivial order 
we obtain (see Sections 4 and 5 in Ref.~\cite{Xue1996}) the composite bosons ($SU_L(2)$-doublet) 
\begin{eqnarray}
H^i=Z_H^{-1/2}(\bar\psi^{ia}_Lt_{Ra}),\quad M^2_H= \frac{4}{N_c}\left(g-\frac{2N_c}{a^2}\right),
\label{boundb}
\end{eqnarray} 
and $t_{Ra}\rightarrow b_{Ra}$, where $Z_H^{1/2}$ and $M_H^2$ are the composite boson form-factor (wave-function renormalization) and mass term. They relate to each other $Z_H^{1/2}\propto M^2_H$ by the Ward identities due to exact symmetries. It is analogous to the relation between the QED wave function and mass renormalizations ($Z_2=Z_1$) due to 
the $U_{em}(1)$ gauge symmetry. The spectra of composite bosons show in Table \ref{qQF0}.
They are complex fields, forming two $SU_L(2)$ doublets 
\begin{eqnarray}
H_{-\frac{1}{2}}=\begin{pmatrix}
H^0_t\\
H^-
\end{pmatrix}\quad {\rm and\rm}\quad  H_{\frac{1}{2}}=\begin{pmatrix}
H^+\\
H^0_b
\end{pmatrix}
\label{doublet}
\end{eqnarray}
of $U_Y(1)$ charge $-1/2$ and $1/2$. The doublet $H_{-\frac{1}{2}}$ ($H_{\frac{1}{2}}$) comes from the top-quark (bottom-quark) channel, namely the first (second) term of four-fermion 
interaction (\ref{rs2}). The composite bosons' masses and form factors are the same at the lowest non-trivial order. The mass degeneracy breaks if we consider the SM family mixing and gauge interactions. They are physical particles, differently from auxiliary static fields $H_0$ (\ref{static}). After the wave function renormalization, the composite bosons behave as elementary particles \footnote{
In this article, we ignore
the composite Dirac fermions: $SU_L(2)$-doublet ${\bf\Psi}^{ib}_D=(\psi^{ib}_L, {\bf\Psi}^{ib}_R)$ and $SU_L(2)$-singlet ${\bf\Psi}^{b}_D=({\bf\Psi}^b_{L},t_R^b)$, where the renormalized composite three-fermion states are:
\begin{equation}
{\bf\Psi}^{ib}_R=(Z_R)^{-1}(\bar\psi^{ia}_Lt_{Ra})t^b_{R}\,;\quad {\bf\Psi}^b_{L}=(Z_L)^{-1}(\bar\psi^{ia}_Lt_{Ra})\psi^{b}_{iL},
\label{bound}
\end{equation}
with mass $M_F=2ga$ and form-factor $Z_{R,L}=M_Fa$, where $Z_{L,R} \approx Z_H^{1/2}Z_\psi^{1/2}$, and 
$Z_\psi^{1/2}$ is the wave renormalization of elementary fermion fields.
Their mass terms $M_F(\bar\Psi_L\psi_R+\bar\psi_L\Psi_R)$ exactly preserve the SM chiral (parity-violating) gauge symmetries. The case for $SU(5)$ chiral gauge symmetry theory is discussed \cite{Eichten1986, Creutz1997}.}.

The one-particle-irreducible (1PI) vertex of the self-interacting term $\lambda(H^\dagger H)^2$ is positive and suppressed by powers $(1/g)^2$. The composite bosons mass term $M^2_H HH^{\dagger}$ (\ref{boundb}) implies 
the second order phase transition of Landau type 
from $M^2_H>0$ to $M^2_H<0$.  $M^2_H=0$  
gives rise to the critical coupling $G^c_{\rm cut}$,
\begin{eqnarray}
G^c_{\rm cut}=2N_ca^2= 2N_c(2\pi)^2 \Lambda^{-2}_{\rm cut},
\label{Gcrit}
\end{eqnarray}
and $G^c_{\rm cut}\Lambda^2_{\rm cut}=2N_c (2\pi)^2\gg 1$.
It is the transition from the 
strong-coupling SM symmetric phase to the weak-coupling SM symmetry breaking phase.

\begin{table*}
%\small\addtolength{\tabcolsep}{-2pt}
\begin{tabular}{ccccc}
Composite $H$ & constituents&  charge $Q_i=Y+t^i_{3L}$ & $SU_L(2)$ 
$t^i_{3L}$& $U_Y(1)$ $Y$\cr
\hline
\hline
$H^+ $& $(\bar b^a_{R}t_{La})$ & $+1$& $~~1/2$ &$1/2$\cr
%\label{qeFf0}\\
$H^-$ & $  (\bar t^a_{R}b_{La})$ &$-1$& $-1/2$& $-1/2$\cr
%\label{qnuFf0}\\
$H^0_t$ & $(\bar t^a_{R}t_{La})$&$ 0$ &$~~1/2$ &$-1/2$\cr
%\label{qeF0}\\
$H^0_b$ & $(\bar b^a_{R}b_{La})$ & $0$ &$-1/2$ &$1/2$\\
\hline
\hline
\comment{
$D^{2/3}_{Rb} $& $b_{Rb}(\bar b^a_{R}t_{La})$ & $2/3$& $~~1/2$ &$1/6$\cr
%\label{qeFf0}\\
$U^{-1/3}_{Rb}$ & $  t_{Rb}(\bar t^a_{R}b_{La})$ &$-1/3$& $-1/2$& $1/6$\cr
%\label{qnuFf0}\\
$D^{-1/3}_{Rb}$ & $b_{Rb}(\bar b^a_{R}b_{La})$ & $-1/3$ &$-1/2$ &$1/6$\cr
%\label{qeF0}\\
$U^{2/3}_{Ra}$ & $t_{Rb}(\bar t^a_{R}t_{La})$&$ 2/3$ &$~~1/2$ &$ 1/6$\\
\hline
\hline
}
\end{tabular}
\caption{Composite bosons and
their constituents carry SM gauge quantum numbers. Composite bosons form two doublets (\ref{doublet}). 
It corresponds to the table V of Ref.~\cite {Leonardi:2018jzn} for the first SM quark family, substituting the $SU_L(2)$ doublet 
$(t_{La}, b_{La})$ into $(u_{La}, d_{La})$ or $(c_{La}, s_{La})$ and singlet 
$t_{Ra}$ into $u_{Ra}$ or $c_{Ra}$, as well as singlet $b_{Ra}$ into $d_{Ra}$ or $s_{Ra}$ in Eq.~(\ref{bhl}). It can also be generalised to the lepto-quark sector, see Eq.~(43) of Ref.~\cite{Xue2017}.}\label{qQF0}
\end{table*}
 
%\vskip0.1cm
\section
{UV scaling domain and effective theory for composite bosons}\label{uvcom}
\hskip0.1cm 
The critical point of second-order phase transition plays a role for a fixed point of field theories \cite{Weinberg1976, ZinnJustin2021, Cardy1997, Brezin1976, 2016pqf..book.....K, Hooft2017}. Calculating the $\beta$-function $\beta=-a(\partial G_{\rm cut}/\partial a)=\mu(\partial G_{\rm cut}/\partial \mu)$, we show \cite{Xue2014} the critical point $G^c_{\rm cut}$ is a UV fixed point 
for the $\beta<0, G_{\rm cut}>G^c_{\rm cut}$ at high energies and $\beta>0,G_{\rm cut}<G^c_{\rm cut}$ at low energies. 

The scaling domain of the UV fixed point $G^c_{\rm cut}(\Lambda)$ renders an SM gauge symmetric effective field theory for composite bosons and fermions at the physical scale $\Lambda$ or the correlation length $\xi=2\pi/\Lambda$ \cite{Xue2014, Xue2017}. The scale $\Lambda\ll \Lambda_{\rm cut}$ or $\xi\gg a$.
Since the critical coupling locates at a strong coupling region, 
it is not an easy task to solve RG equations. However, we know 
that in the neighbourhood of the critical point 
$G^c_{\rm cut}$, the correlation length $\xi/a$ 
of the theory goes to infinity, leading to the scaling invariance. 
Namely, in the UV scaling domain, renormalised masses and form-factors of composite particles obey the RG equations and are independent of the UV cutoff $\Lambda_{\rm cut}=2\pi/a$. Using such property in the UV scaling domain, one expands the running coupling $G_{\rm cut}(a/\xi)$ as a series,
\begin{eqnarray}
G_{\rm cut}(a/\xi)&=& G^c_{\rm cut} \left[1+ c_0(a/\xi)^{1/\nu}+ {\mathcal O}[(a/\xi)^{2/\nu}]\right]\rightarrow G^c_{\rm cut}+0^+,
\label{gexp}
\end{eqnarray}
for $a/\xi\ll 1$, corresponding to the expansion of $\beta$-function as a series, 
\begin{eqnarray}
\beta(G_{\rm cut})&=& (-1/\nu) (G_{\rm cut}-G^c_{\rm cut}) + {\mathcal O}[(G_{\rm cut}-G^c_{\rm cut})^2]<0\,.
\label{betaexp}
\end{eqnarray}
The correlation length $\xi$ follows the scaling law
\begin{eqnarray}
\xi &= & c_0 a\exp \int^{G_{\rm cut}}\frac{dG'_{\rm cut}}{\beta(G'_{\rm cut})}
=\frac{ (c_0G^c_{\rm cut})^\nu a}{(G_{\rm cut}-G^c_{\rm cut})^\nu},
\label{xivary}
\end{eqnarray}
where the coefficient $(c_0G^c_{\rm cut})^\nu$ and critical exponent $\nu$ need to be determined by
non-perturbative numerical simulations. 

Analogously to the electroweak scale 
$v$ sets into the scaling region of the IR-stable fixed point $G_c$ (the IR scaling domain),
the physical scale $\Lambda$ sets into the scaling 
region of the UV-stable fixed point $G^c_{\rm cut}$ (the UV scaling domain).
Equations (\ref{gexp}) and (\ref{betaexp}) imply the running coupling  
$G(\mu)|_{\mu\rightarrow\Lambda +0^+}\rightarrow G^c_{\rm cut}$,
\begin{eqnarray}
G(\mu)\approx G^c_{\rm cut}\left[1-\frac{1}{\nu}
\ln\frac{\mu}{\Lambda}\right]^{-1}> G^c_{\rm cut}, \quad \frac{\mu}{\Lambda}=\frac{\xi}{(ac_0^\nu)} \gtrsim 1,
\label{xivary1}
\end{eqnarray}
in the UV scaling domain. The running scale $\mu\gtrsim\Lambda$ physically indicates the energy transfer between constituents inside composite particles. The scaling law (\ref{xivary}) becomes
\begin{eqnarray}
a \approx   \left(\frac{2\pi}{\Lambda }\right)\left[\frac{\frac{1}{\nu}
\ln\frac{\mu}{\Lambda}}{c_0\left(1-\frac{1}{\nu}
\ln\frac{\mu}{\Lambda}\right)}\right]^\nu .
\label{runnc}
\end{eqnarray}
All 1PI functions $\Gamma[\mu, G(\mu)]$ 
of the quantum field theory 
(\ref{bhl}) 
evolve to relevant $(d=4)$ or irrelevant $(d>4)$ 1PI functions, as the energy scale $\mu$ varies. 

Using the critical coupling 
$G_{\rm cut}^c$ (\ref{Gcrit}) and 
scaling law (\ref{runnc}) in the UV domain, we obtain from (\ref{boundb}) the composite 
boson running masses 
\begin{eqnarray}
\comment{
M_F &\approx& 4N_c c_0^\nu\left(\frac{\Lambda }{2\pi}\right) \frac{\left[1-\frac{1}{\nu}
\ln\left(\frac{\mu}{\Lambda}\right)\right]^{\nu -1}}{\left[\frac{1}{\nu}
\ln\left(\frac{\mu}{\Lambda}\right)\right]^{\nu}}\rightarrow 4N_c c_0^\nu\left(\frac{\Lambda }{2\pi}\right) \left[\frac{1}{\nu}
\ln\left(\frac{\mu}{\Lambda}\right)\right]^{-\nu},\label{cfm}\\
}
M_{H}(\mu) &\approx& 2\sqrt{2} c_0^\nu\left(\frac{\Lambda }{2\pi}\right) \frac{\left[1-\frac{1}{\nu}
\ln\left(\frac{\mu}{\Lambda}\right)\right]^{\nu -1}}{\left[\frac{1}{\nu}
\ln\left(\frac{\mu}{\Lambda}\right)\right]^{\nu-1}},
\label{chm}
\end{eqnarray}
and form factor $Z_H^{1/2}(\mu)\propto M^2_{H}(\mu)$. They do not depend on the cutoff scale $\Lambda_{\rm cut}$ (\ref{bhl}).
The values $M_H=M_H(\mu)|_{\mu=M_H}$ and $Z_H^{1/2}=Z_H^{1/2}(\mu)|_{\mu=M_H}$ on the mass shell are in the order of the scale $\Lambda$. They need to be determined by experiments. 
%and their ratio $M_{H}/M_F=\frac{1}{\nu}\ln\left(\frac{\mu}{\Lambda}\right)/(\sqrt{2}N_c)$. 

In the UV scaling domain, the SM gauge symmetric Lagrangian of massive composite boson doublets (\ref{doublet}) at the energy scale $\mu> \Lambda$ is given by
\begin{eqnarray}
L_H &=& L_{\rm kinetic} + g_H(\bar\psi_L t_RH_{-1/2}+\bar\psi_L b_RH_{1/2})+ {\rm h.c.}
\nonumber\\ 
&-&\frac{1}{2}M_{H}^2H^2
-\frac{\lambda_H}{2}(H^2)^2,
\label{effH}
\end{eqnarray}
where $H^2=H^\dagger H=H^\dagger_{-1/2}H_{-1/2}+H^\dagger_{1/2}H_{1/2}$. The Yukawa coupling and mass terms are respectively
\begin{eqnarray}
 g_H(\bar b_L t_RH^- + \bar t_L t_RH^0_t +  \bar b_L b_RH^0_b + {\rm h.c.}),
\label{youkawa}\\
\frac{1}{2}M_{H}^2\left[(H^0_t)^2+(H^0_b)^2+2H^+H^-)\right],
\label{effmass}
\end{eqnarray}
in terms of composite bosons (Table \ref{qQF0}).
The Yukawa coupling $g_H=(F_H/\Lambda)^2\sim {\mathcal O}(1)$ and the decay constant $F_H\propto Z_H^{1/4}\sim \Lambda$.
In the kinetic term $L_{\rm kinetic}$, the covariant derivative $|D_\mu H|^2$ couples the composite bosons $H$ to the SM gauge bosons $\gamma$, $W^\pm$, $Z^0$ and gluons. The charged $W^\pm$, neutral 
$Z^0$ and $A$ gauge bosons' couplings to the doublets $H_{\pm \frac{1}{2}}$ (\ref{doublet}) can be derived from their covariant derivatives $D_\mu H_{\pm \frac{1}{2}}$, see Eqs.~(45-48) of Ref.~\cite{Xue2017}. All gauge bosons and elementary fermions $(t,b)$ are massless due to the exact SM chiral gauge symmetries and associated Ward identities. The renormalised Yukawa coupling $g_H$, boson mass $M_{H}$, and quartic coupling $\lambda_H$ represent the $d=4$ relevant operators in the UV scaling domain. Namely, the boson mass $M_{H}$, Yukawa coupling $g_H$ and quartic coupling $\lambda_H$ depend only on a logarithmic function $\ln(\mu/\Lambda)^2$, independently of the cutoff $\Lambda_{\rm cut}=2\pi/a$ (\ref{bhl}). We treat positive couplings $g_H$ and $\lambda_H$ as free parameters in the effective Lagrangian (\ref{effH}).

Compared with the conventional assignment of one SM elementary Higgs doublet $\Phi=\begin{pmatrix}
\phi^+\\
\phi^0
\end{pmatrix}$, $\tilde\Phi=i\tau_2\Phi^*$ and two Yukawa 
couplings $f^{(u)}\bar q _L \tilde \Phi u_R+f^{(d)}\bar q _L \Phi d_R$ (see for example (11.62) in Ref.~\cite{Cheng1984a}),
we figure out the correspondence  
$H_{-1/2}\leftrightarrow \tilde\Phi$ up to a phase for the top-quark channel. While for the bottom-quark channel, $H_{1/2}$ represents another composite boson doublet of the effective theory (\ref{effH}).

We briefly compare and contrast our model with the manifest left-right symmetry (LR) model \cite{Beg1977, Chakrabortty2012}. 
The similarity between the two models is chiral gauge symmetric at high energies. 
The main differences are in gauge symmetries and bosons. The LR model introduces the new gauge symmetry $SU_R(2)$ and extra gauge bosons, e.g.~$W_R$, as well as an additional Higgs sector. Our model has only the SM gauge symmetries and gauge bosons $W$ and $Z$.

%\vskip0.1cm
\section
{First-order phase transition}\label{firstorder}
\hskip0.1cm 
As the running energy scale $\mu$ decreases, the second-order continuous phase transition of spontaneous symmetry breaking (SSB) is supposed to occur at the critical scale $\mu_{\rm crit}=e^\nu \Lambda> \Lambda$, when $M^2_H=0$ (\ref{chm}).
In this section, we study some details about %make scrutiny into 
the phase transition from the symmetric phase for the effective composite boson theory in the UV scaling domain to the symmetry-breaking phase for the SM in the IR scaling domain. 

In the Lagrangian (\ref{effH}) in the UV scaling domain, we integrate the elementary fermion fields over their quadratic fermion action $iS_{\rm ferm}(\bar\psi,\psi,H)$ by standard nonperturbative 
methods \cite{Berezin1966,Itzykson:1980rh} to obtain
%(after Wick rotation to the Euclidean spacetime), https://web2.ph.utexas.edu/~vadim/Classes/2022f/FPI.pdf
%https://web2.ph.utexas.edu/~vadim/Classes/2022f/notes.html
%https://arxiv.org/pdf/hep-th/9910186.pdf
the effective action $iS_{\rm eff}(H)$,
\begin{eqnarray}
\exp iS_{\rm eff} &=& \int d\bar\psi d\psi \exp iS_{\rm ferm}(\bar\psi,\psi,H)\Rightarrow \nonumber\\
%\exp - \int_x d^4x \bar\psi {\mathcal O}\psi = \det {\mathcal O}\nonumber\\
S_{\rm eff}(H^2)  &=& - \frac{1}{2} \int^\Lambda\frac{d^4p}{(2\pi)^4} 
\ln [p^2+g^2_HH^2],
\label{func} 
\end{eqnarray}
where Euclidean four momenta $p$ of elementary fermions are integrated up to the scale $\Lambda$ of the UV scaling domain. The reason is that the energy momenta of elementary fermion fields are regularised by the scale $\Lambda$ of composite particles' dynamics. The integral receives dominant contributions from the vacuum quantum fluctuations 
(fermion loops) at the scale $\Lambda$. %For these high-energy fermion modes, we approximately neglect their masses. 
Equation (\ref{func}) yields,  
\begin{eqnarray}
-S_{\rm eff}(H^2) 
&=& \frac{g^2_H}{2^6\pi^2}\left\{H^2\Lambda^2 -g^2_H(H^2)^2\ln (1+\frac{\Lambda^2}{g^2_HH^2})+\Lambda^4\ln [g^2_HH^2+\Lambda^2]\right\}\nonumber\\
&\approx & \frac{g^2_H}{2^6\pi^2}\left\{H^2\Lambda^2 +g^2_H(H^2)^2\ln (\frac{g^2_HH^2}{\Lambda^2})\right\},
\label{vfunc0} 
\end{eqnarray}
where for $\Lambda^2 \gg H^2$ we make the
Taylor expansion in powers of $(H^2/\Lambda^2)$ up to ${\mathcal O} (H^2/\Lambda^2)^2$. We ignore the terms %$-\Lambda^4/2$ 
that are dynamically independent of the composite fields $H$ since they are overall constants.

As a result, we obtain the effective potential of composite bosons of the Lagrangian (\ref{effH}) in the UV scaling domain, 
\begin{eqnarray}
V_{\rm eff}(H^2) &\approx & \frac{1}{2}M_H^2H^2
+\frac{\lambda_H}{2}(H^2)^2 + S_{\rm eff}(H^2)\nonumber\\
&=& \frac{1}{2}\bar M_H^2H^2
+\frac{\lambda_H}{2}(H^2)^2  +\frac{g^4_H(H^2)^2}{2^6\pi^2}\ln (\frac{g^2_HH^2}{\Lambda^2}), 
\label{vfunc} 
\end{eqnarray}
where the corrected mass term
\begin{eqnarray}
\bar M_H^2= M_H^2
+\frac{g^2_H}{2^6\pi^2}\Lambda^2. 
\label{cmass} 
\end{eqnarray}
%The effective potential is similar but not identical to the one obtained in studying one-loop radiative corrections to the Higgs scalar-field potential \cite{Coleman1973, Weinberg1973}. 
Its first and second derivatives w.r.t.~the filed $H$ are
\begin{eqnarray}
V'_{\rm eff}(H^2) &\approx & \bar M_H^2H
+2(\lambda_H+\frac{g^4_H}{2^6\pi^2}) H^3 +4\frac{g^4_HH^3}{2^6\pi^2}\ln (\frac{g^2_HH^2}{\Lambda^2}),
\label{dv1}\\
V''_{\rm eff}(H^2) &\approx & \bar M_H^2
+(6\lambda_H+14\frac{g^4_H}{2^6\pi^2}) H^2 +12\frac{g^4_H}{2^6\pi^2}H^2\ln (\frac{g^2_HH^2}{\Lambda^2}).
\label{dv2}
\end{eqnarray}
Equations (\ref{vfunc}-\ref{dv2}) provide an effective potential approach to qualitatively study the symmetric ground state of composite boson theory (\ref{effH}) and its transition to the SSB ground state.    

\begin{figure*}[t]
%\begin{center}
\hskip0.9cm\includegraphics[height=2.8in]{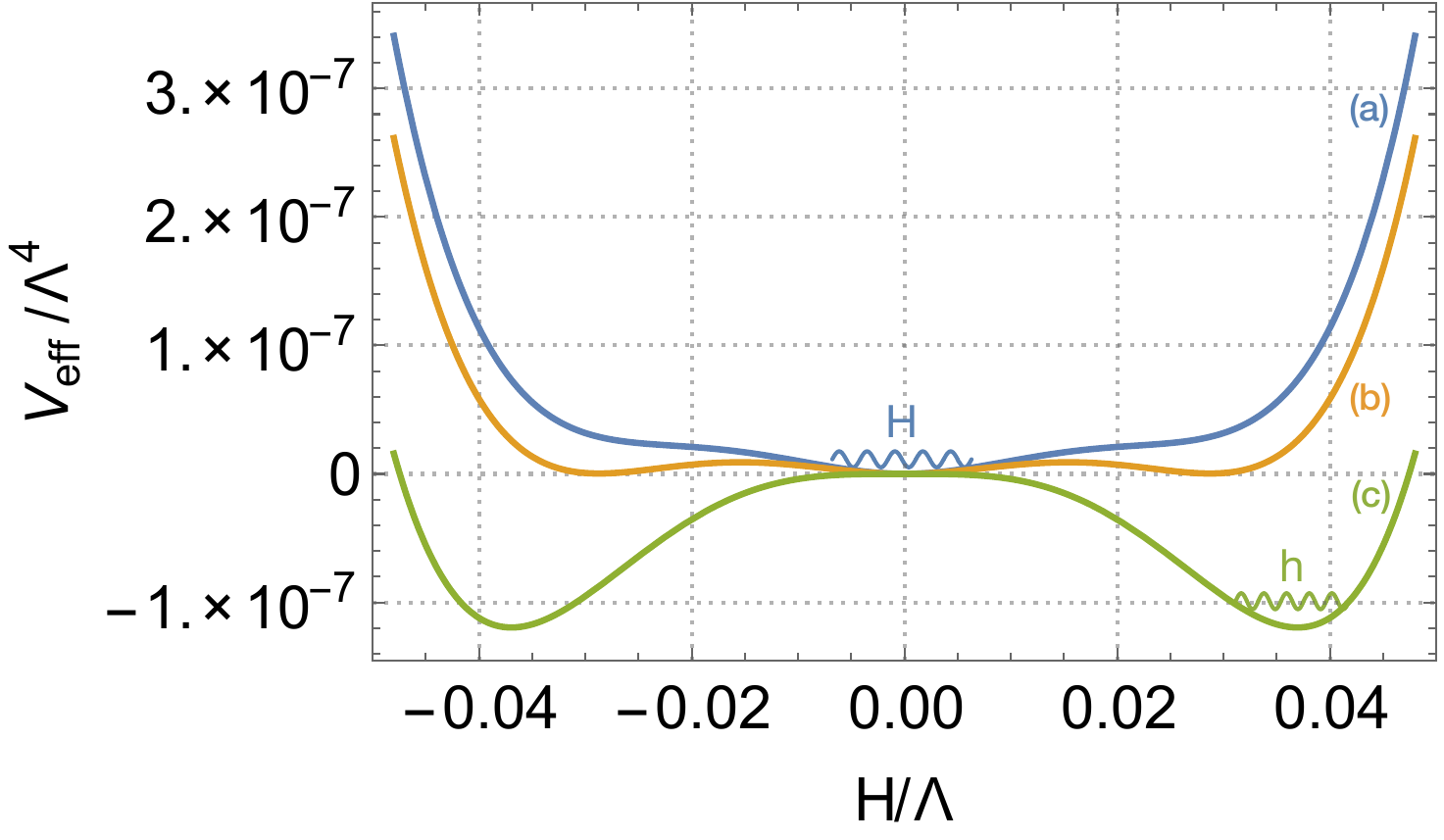}
%\vskip-1.1in
%\put(-115,175){$\tilde\lambda(\mu)$}
%\put(-305,175){$\bar g_t(\mu)$}
\caption{The effective potential (\ref{vfunc}) for different values of the mass term $(\bar M^2_{H})$. \blue{(a)} The blue line $(\bar M^2_{H})\approx 0.0168$. \textcolor{orange}{(b)} The orange line 
$(\bar M^2_{H})\approx 0.0146$. \green{(c)} The green line $(\bar M^2_{H})\gtrsim 0.0$. In all these cases, the effective potential has a positive curvature $V''_{\rm eff}=(\bar M^2_{H})^2>0$ at $H=0$, indicating a weak first-order phase transition. It differs from the second order phase transition occurring when $V''_{\rm eff}=0$ at $H=0$. The blue $H$ wave line indicates the composite boson modes upon the symmetric ground state. The green $h$ wave line indicates the $\langle\bar tt\rangle$-composite Higgs boson modes upon the symmetry-breaking ground state. The values $g_H=3.0$ and 
$\lambda_H=1.0$ are used.} 
\label{figV}
%\end{center}
\end{figure*}

\begin{figure*}[t]
%\begin{center}
\hskip1.30cm 
\includegraphics[height=2.8in]{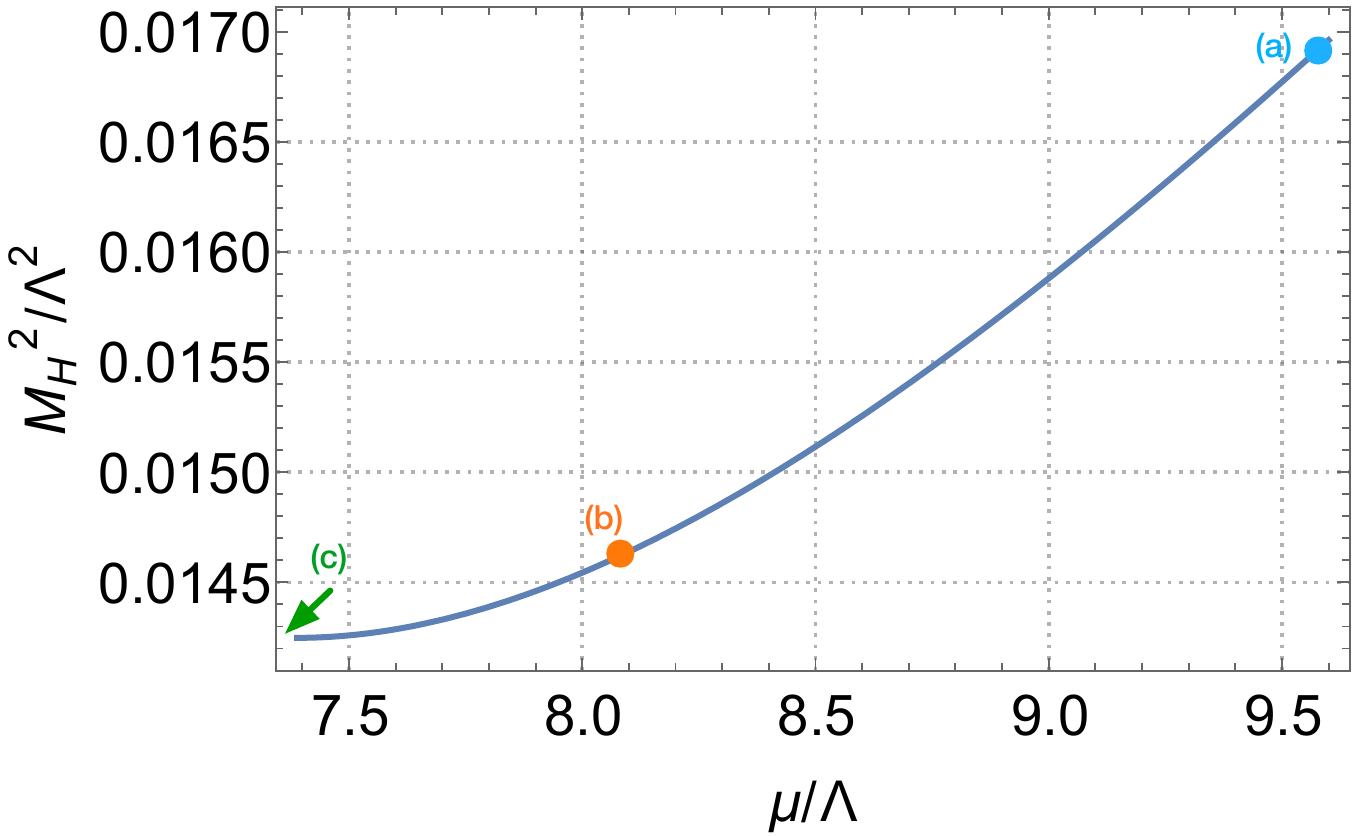}
%\vskip-1.1in
%\put(-115,175){$\tilde\lambda(\mu)$}
%\put(-305,175){$\bar g_t(\mu)$}
\caption{We plot the mass term (\ref{chm}) of composite bosons as a function of running energy scale $\mu$ in the range where the SSB possibly occurs. Corresponding to three lines in Fig.~\ref{figV},
we indicate values \blue{(a)} $(\bar M^2_{H})\approx 0.0168$ for the blue spot; \textcolor{orange}{(b)} $(\bar M^2_{H})\approx 0.0146$ for the orange spot; \green{(c)} $(\bar M^2_{H})\gtrsim 0.0$ for the green arrow. The $\bar M^2_{H}\not=0$ implies a weak first-order phase transition. The values $\nu=2.0$ and 
$c_0=1.0$ are used.} 
\label{figM}
%\end{center}
\end{figure*}

Given an energy scale $\mu>\Lambda$ in Eq.~(\ref{chm}) for which $\bar M^2_{H}(\mu)>0$, the potential $V_{\rm eff}(H^2)$ has a minimum $H=0$ and $V_{\rm eff}(0)=0$, see the blue line (a) in 
Fig.~\ref{figV}. It represents the symmetric ground state of the effective theory (\ref{effH}) of composite fields $H$ in the UV scaling domain. As quantum excitation upon this ground state, the composite bosons $H_{-\frac{1}{2}}$ and $H_{\frac{1}{2}}$ (see Table \ref{qQF0}) are physical particles of masses $\bar M^2_{H}(\bar M_{H})>0$, interacting to SM elementary fermions and gauge bosons.  

Figure \ref{figM} shows the mass term $\bar M^2_H(\mu)$ (\ref{cmass}) of composite bosons as a function of running scale $\mu/\Lambda$. The mass term $\bar M^2_H(\mu)$ decreases as $\mu$ decreases, and the effective potential $V_{\rm eff}$ develops the second minimum. At a positive 
critical mass $\bar M^2_{H}=(\bar M^2_{H})_c$, the double zero minima appear, 
the second minimum $V(H_m)=0$ locates at $H_m\not=0$, see the orange line (b) in 
Fig.~\ref{figM}. 
Its value is determined by $V_{\rm eff}(H^2_c)=0$ and $V'_{\rm eff}(H^2_c)=0$,
\begin{eqnarray}
(\bar M^2_{H})_c&=& 2\frac{g_H^4}{2^6\pi^2} H^2_c=2\frac{g_H^2\Lambda^2}{2^6\pi^2}\exp - \frac{1}{2} \left(\frac{2^6\pi^2}{g_H^4}\lambda_H+2\right),
\label{crim}\\
H^2_c&=& \Lambda^2 g_H^{-2}\exp - \frac{1}{2} \left(\frac{2^6\pi^2}{g_H^4}\lambda_H+2
\right),
\label{crih}
\end{eqnarray}
corresponding to the critical scale $\mu_{\rm crit}\gtrsim e^\nu\Lambda$. 
We define (\ref{crim}) and (\ref{crih}) as the critical point for the SSB phase transition taking place because the minimum $V_{\rm eff}(H^2_c)$ becomes negative and lower than the minimum $V_{\rm eff}(H^2=0)$ of the symmetric phase. The composite bosons $H$ upon the $H=0$ ground state become quantum-field fluctuating unstable. The critical scales $(M_{H})_c$ and $H_c$ 
are small compared with the scale 
$\Lambda$. We emphasise that such critical point (\ref{crim}) and (\ref{crih}) {\it must} exist, as long as couplings $g_H$ and $\lambda_H$ are finite values. 

If it were not for the last term $ g^4_H(H^2)^2\ln (\Lambda^2/g^2_HH^2)$, 
Equation (\ref{vfunc}) would be of the
Landau-type and the phase transition would be  one of the second order, continuously {\it slipping}
from the symmetric ground state $H=0$ and $\bar M_H^2> 0$ to the SSB ground state $H\not=0$ and $\bar M_H^2 <0$. 
The logarithmic term destroys
the Landau expansion and second-order phase transition. At the critical point (\ref{crim},\ref{crih}) for the SSB phase transition, we find slightly positive curvature $(\bar M_H^2)_c$ at $H$=0. Therefore, it is  {\it 
a weak first-order phase transition}, because
the fields $H$ {\it jump} from the 
zero expectation value $\langle H\rangle=0$ ground state of the symmetric phase to the nonzero expectation value $\langle H\rangle\not=0$ ground state of the symmetry-breaking phase. However, one can only use this argument qualitatively.  
This situation is similar to 
%consistent with the suggestion 
the one studied in radiative corrections to the Higgs scalar-field 
potential \cite{Coleman1973,Weinberg1973}, the Gaussian effective potential \cite{Barnes1980,Stevenson1985}, the pure $\lambda\phi^4$ theory \cite{Consoli2000,Consoli2020}
and the phase transition of the system with finite matter density \cite{Kleinert2015}. 
%Note that the argument can only be used qualitatively, as been emphasized by Colemen and Weinberg \cite{Coleman1973,Weinberg1973}. The reason is that the new minimum $H^2_m$ lies in a field regime where the perturbation theory is no longer reliable \cite{Fiolhais2013,Kleinert2015}.

%\vskip0.1cm
\section
{Weak coupling gauge-symmetry-breaking phase}\label{weakc}
\hskip0.1cm
We turn to study the symmetry-breaking phase in the weak coupling regime, where the SSB dynamics undergo, in contrast with the strong-coupling symmetry phase, where the composite dynamics occur. %, see Sec.~\ref{scsp}. 
%It is important to point out that the ground state of SSB phase can occur for any positive value $(\bar M^2_{H})$ in the range from zero to $(\bar M^2_{H})_c>0$. {\it A priori}, we cannot determine where it occurs.
Observe the smallness of the critical mass $(\bar M^2_{H})_c\gtrsim 0$. For illustration, we consider the case $\bar M^2_{H}\approx 0$ for studying the SSB ground state.
%\footnote{Colemen and Weinberg suggest for the occurrence of SSB\cite{Coleman1973, Weinberg1973}}.
Note that $\bar M_H^2\approx 0$ gives a critical scale, indicated as (c) in Fig.~\ref{figM}, which is smaller than the critical scale $\mu_{\rm crit}$ obtained by  $\bar M^2_{H}=(\bar M^2_{H})_c$ (\ref{crim}), indicated as (b) in Fig.~\ref{figM}.
The green line (c) in Fig.~\ref{figV} shows 
that for the case $\bar M^2_{H}\approx 0$, the effective potential (\ref{vfunc}) has a minimum
\begin{eqnarray}
H^2_m= \Lambda^2 g_H^{-2}\exp - \frac{1}{2} \left(\frac{2^6\pi^2}{g_H^4}\lambda_H+1
\right)=e^{1/2}H^2_c,
\label{mim} 
\end{eqnarray}
$V'_{\rm eff}(H^2_m)=0$, at which positive curvature 
\begin{eqnarray}
m^2_H=V''_m(H^2_m) = \frac{8g_H^4}{2^6\pi^2} H^2_m \ll \Lambda^2 ,
\label{minc} 
\end{eqnarray}
and the minimal value of the field potential is
\begin{eqnarray}
V_m(H^2_m) = -\frac{1}{2} \frac{g_H^4}{2^6\pi^2} (H^2_m)^2<0.
\label{minv} 
\end{eqnarray}
The nonzero expectation value $H_m$ (\ref{mim}) should account for the scale of the SSB ground state. The $m^2_H$ (\ref{minc}) should relate to the mass scale of quantum fluctuating fields $h=H-H_m$ upon the SSB ground state. The symmetry-breaking scale $H_m$ is smaller than the scale $\Lambda$ of the UV scaling domain, and the mass scale $m_H$ is smaller than the composite boson masses $M_H$ (\ref{chm}).
Except for the mass scales $m_H< M_H$, the fields $h$ of Higgs type are different from the composite boson fields $H$ of Table \ref{qQF0} on the ground state of zero expectation value $H=0$ 
in the symmetric phase. The quantum fields $h$ and $H$ may have the same quantum numbers of SM gauge interactions, for instance, $\langle\bar tt\rangle$-composite Higgs boson and $H_t^0\sim (\bar t_Rt_L)$ composite boson. 
  
However, we cannot consider the symmetry-breaking scale $H_m$ (\ref{mim}) as the SM vacuum expectation value $v=246$ GeV and the mass $m_H$ (\ref{minc}) as the observed Higgs mass $m_h=126$ GeV. 
We need to answer the following two questions about the dynamics proceeding during and after the weak first-order phase transition at the scale $\Lambda$. The first one is why the SSB dynamics yields only one heaviest fermion mass in the top-quark channel, together with a $\langle\bar tt\rangle$-composite Higgs boson and three composite Goldstone bosons $\langle\bar t\gamma_5t\rangle$, $\langle\bar b\gamma_5t\rangle$ and $\langle\bar t\gamma_5b\rangle$ for the longitudinal modes of massive gauge bosons $W^\pm$ and $Z^0$. How the SSB dynamics evolves with energy scale $\mu$ from $\Lambda$ to $v$, 
approaching the RG equations in the IR scaling domain 
for the low-energy SM effective 
Lagrangian \cite{Bardeen1990}. The second,  composite bosons in Table \ref{qQF0} 
are massive and tightly bound. How the composite bosons in the top-quark channel become loosely bound Higgs and Goldstone bosons aforementioned. Whereas composite bosons in other channels, e.g.~$H_b^0\sim (\bar b_Rb_L)$ of the bottom channel, dissociate into their constituents of SM fermions.
The same questions are for the 
dynamics as the energy scale $\mu$ increases from $v$ to $\Lambda$. How elementary fermions form composite particles, how the SM  evolves into the effective theory of composite particles at the scale $\Lambda$ and the chiral gauge symmetries restore.

%\vskip0.1cm
\section
{Dissociation phenomenon in first-order phase transition}\label{disso}
\hskip0.1cm 

We present some discussions on the dissociation phenomenon of composite bosons dissolving to their constituents of SM elementary fermions. It leads to a Higgs boson and three Goldstone bosons in the first-order transition from the symmetric phase to the SSB phase.      

The strong four-fermion coupling $G_{\rm cut}$ dynamics (\ref{rs2}) form composite bosons, whose propagators have 
poles and residues that respectively 
represent their mass terms $M^2_H$ and form-factors $Z_H^{1/2}$ (\ref{boundb}).  
As long as their form factors are finite, after wave-function renormalizations, these composite particles behave as elementary particles in the UV scaling domain at the scale $\Lambda$, when the energy scale $\mu$
decreases to the critical scale (threshold) $\mu_{\rm crit}$ and $G(\mu)\rightarrow 
G^c_{\rm cut}(\mu_{\rm crit})+0^+$, composite particles' form-factors and binding energy vanish. 
All three-fermion and two-fermion bound states (poles) dissolve into their constituents, which represent three-fermion and two-fermion 
cuts in the energy-momentum plane. This is the Weinberg %-Salam 
compositeness condition \cite{Weinberg1963,Weinberg1963a,Weinberg1964,Weinberg1965}. It can also be viewed from the orange line (double zeros) in Fig.~\ref{figV} when the effective potential is in the critical phase transition. Composite bosons $H_{\pm\frac{1}{2}}$ or $H^{0,\pm}$ become unstable due to quantum-field fluctuations. 
The critical scale $\mu_{\rm crit}>\Lambda$ and mass $(\bar M_H)_c < \Lambda$ are given in Eq.~(\ref{crim}).

In the UV scaling domain, composite bosons' propagators give their poles, i.e., mass-shell conditions
\begin{eqnarray}
E_{\rm com}=\sqrt{p^2+M_H^2}\approx M_H,\quad {\rm for }~~p\ll M_H.
\label{Hshell}
\end{eqnarray}
The mass term $M_H$ contains the negative binding energy 
${\mathcal B}[G_{\rm cut}(\mu)]$ and positive kinetic energies ${\mathcal K}$ of composite-boson constituents. The binding energy
is attributed to the strong attractive coupling $G_{\rm cut}(\mu)> G^c_{\rm cut}(\mu_{\rm crit})$ and $\mu>\mu_{\rm crit}>\Lambda$. The positive kinetic energies ${\mathcal K}$ represent the relative energy transfer $\mu=\bar q-q$ between fermion and antifermion inside composite bosons. 
When the energy scale $\mu$ approaches to the critical energy scale $\mu_{\rm crit}\gtrsim e^\nu \Lambda$, the composite boson mass term $M_H^2$ approaches to $(M_H)^2_c$ (\ref{crim}). It means that the binding energy ${\mathcal B}$ is equal to the kinetic energies ${\mathcal K}$,  
${\mathcal B}[G(\mu)]_{\mu\rightarrow\mu_{\rm crit}}\rightarrow \mu_{\rm crit}$ \footnote{In previous publications, we use the threshold energy scale ${\mathcal E}_{\rm thre}$ for the critical scale $\mu_{\rm crit}$}. 
Therefore, the composite bosons are critical binding for their mass gaps vanishing $(\bar M_H^2)_c\rightarrow 0$ and spatial size increasing (form factor $Z_H^{1/2}\rightarrow 0$). 
As a result, they become unstable because of quantum-field fluctuations leading to their dissolving into SM elementary fermions. Such dissociation phenomenon is associated with the first-order phase transition that occurs from the 
symmetric phase to the symmetry-breaking phase.
Its reversed process is the combination phenomenon as increasing energy scale $\mu$ approaches the critical value $\mu_{\rm crit}+0^-$, and the first-order phase transition occurs from the 
symmetry-breaking phase to the symmetric phase. The detailed study of the phenomenon and dynamics is complex. For example, a molecular hydrogen gas system forms spontaneously from a neutral system of protons and electrons in the first-order phase transition \cite{Magro1996}.

Another question is why the SSB ground state selects the top quark as the most massive fermion, accompanied by a Higgs and three Goldstone bosons. 
Figure \ref{figV} shows the SSB ground state $H_m\not=0$ is energetically favourable, compared with symmetric one $H=0$. 
When spontaneously generated fermion mass 
develops a negative mass gap and contributes to the vacuum energy. We can see from Eq.~(\ref{func}) such property that the fermionic loop contribution to the vacuum energy is negative. In contrast, the bosonic loop contribution to the vacuum energy is positive. 
%See for example \cite{Xue2013c}.
%https://arxiv.org/pdf/1301.4254.pdf
As a result, the SSB ground state establishes by minimising the total energy, and only the top quark acquires its mass with the least number of Goldstone bosons \cite{Xue2013c, Preparata1996}.
It gives the reason why the nonzero expectation value $H_m=\langle H\rangle$ of the SSB ground state is developed only in the top-quark channel of four-fermion interaction (\ref{rs2}), namely the top-condensate model \cite{Bardeen1990}. 
On such SSB ground state, the composite doublet $H_{-\frac{1}{2}}$ becomes
\begin{eqnarray}
H_{-\frac{1}{2}}=\begin{pmatrix}
H^0_t\\
H^-
\end{pmatrix}\Rightarrow \exp [i\vec\zeta(x)\cdot \vec\tau]\begin{pmatrix}
\frac{v+h(x)}{\sqrt{2}}\\
0
\end{pmatrix},
\label{HiggH}
\end{eqnarray}
where $\vec \zeta(x)$ represents three Goldstone bosons ($\vec \tau$ are Pauli matrices) from the pseudo-scalar components of composite fields $H^{\pm,0}$. 
The Higgs-like boson $h(x)$ is a $\langle\bar t t\rangle$-bound state as if an elementary Higgs boson field.  Whereas another composite boson $H_b^0$ in the doublet $H_{\frac{1}{2}}$ from the bottom channel of four-fermion interaction (\ref{rs2}) dissociates to its fermionic constituents, as previously discussed.

Based on these discussions, we advocate the 
scale hierarchy  
\begin{eqnarray}
v\ll \Lambda \lesssim \mu_{\rm crit}\ll \Lambda_{\rm cut}, %\quad \Lambda\approx 5\, {\rm TeV}.
\label{scales}
\end{eqnarray}
in the scenario. (i) At the cutoff scale $\Lambda_{\rm cut}$, four-fermion operators (\ref{bhl}) of Einstein-Cartan or NJL type are present due to quantum gravity or new physics. (ii) At the scale $\Lambda$, the effective field theory of SM gauge symmetric Lagrangian for composite particles (\ref{effH}) realizes in the UV scaling domain of strong four-fermion coupling $G_{\rm cut}^c$ (\ref{Gcrit}). (iii) The weak first-order phase transition of SSB and dissociation occurs at the critical scale
$\mu_{\rm crit}$. It leads to the effective field theory of SM gauge symmetry-breaking Lagrangian at the electroweak scale $v$ for the massive top quark and $\langle\bar tt\rangle$-composite Higgs boson, etc., which realizes in the IR scaling domain of weak four-fermion coupling $G_c$, We will discuss it in the next section.

%\vskip0.1cm
\section{IR scaling domain and effective theory for SM }\label{SSBS}
\hskip0.1cm 

We attempt to discuss the connection between the 
effective theory of composite particles' Lagrangian and the effective theory of the 
Lagrangian for the SM. 
In the effective Lagrangian (\ref{effH}), the composite fields $H$ are massive ($M_H\propto\Lambda$). 
Thus, we treat $H$ as static fields by 
approximately neglecting their kinetic and 
quartic terms, 
\begin{eqnarray}
L_H\Rightarrow g_H(\bar \psi_L t_RH_{-\frac{1}{2}}+ \bar \psi_L b_RH_{\frac{1}{2}} + {\rm h.c.}) 
- \frac{1}{2}M_H^2 (H^\dagger_{-1/2}H_{-1/2}+H^\dagger_{1/2}H_{1/2}).
\label{staticH}
\end{eqnarray}
Integrating over ``static'' fields $H_{-1/2}$ and $H_{1/2}$, we obtain the effective four-fermion operators
\begin{eqnarray}
G\Big[(\bar\psi^{ia}_Lt_{Ra})(\bar t^b_{R}\psi_{Lib}) + (\bar\psi^{ia}_Lb_{Ra})(\bar b^b_{R}\psi_{Lib})\Big],
\label{bhllow}
\end{eqnarray}
where the coupling $G=2g_H/M^2_H\propto \Lambda^{-2}$ at the scale $\Lambda$. It means that in the gauge symmetric UV scaling domain, the massive composite boson Lagrangian (\ref{effH}) or (\ref{staticH}) is approximately equivalent to the four-fermion operators (\ref{bhllow}). Since the SM gauge symmetries preserve, the interaction (\ref{bhllow}) has the same structure as the four-fermion operators (\ref{bhl}) at the cutoff scale $\Lambda_{\rm cut}$. But the effective couplings are at different scales $\Lambda\ll \Lambda_{\rm cut}$, the former (\ref{bhl}) is $G_{\rm cut} \propto \Lambda^{-2}_{\rm cut}$ and 
the latter (\ref{bhllow}) is $G \propto \Lambda^{-2}$.

\subsection{Effective SM Lagrangian and RG equations}

Based on the first term ($\langle\bar tt\rangle$ channel) of interactions (\ref{bhl}),
% for weak coupling $G\ll G^c_{\rm cut}$ (\ref{Gcrit}), 
the Reference \cite{Bardeen1990} presents the detailed investigations on the SSB dynamics of $\langle\bar tt\rangle$ condensation,
the IR fixed point $G_c=8\pi^2/(N_c\Lambda^2_{\rm cut})$, 
and the effective Lagrangian in the IR scaling domain $G\gtrsim G_c$ at the scale $v$. It 
describes the SM of the massive top quark, gauge bosons ($W^\pm, Z^0$) and $\langle\bar tt\rangle$-composite Higgs boson. Three composite Goldstone modes become $W^\pm$ and $Z^0$ longitudinal components.
The effective SM Lagrangian with the {\it bilinear} top-quark mass term and Yukawa coupling to the composite Higgs boson $h$ at the low-energy
scale $\mu$ reads \footnote{We have generalised the BHL approach to the right-handed neutrino sector for discussing dark matter particles \cite{Xue2022a}.}
\begin{eqnarray}
L &=&  L_{\rm kinetic} 
+ \Delta L_{\rm gauge} + \Delta L_{\rm irr},
\nonumber\\ 
&+& \bar g_{t}(\bar \psi_L t_Rh+ {\rm h.c.})+|D_\mu h|^2-m_{h}^2h^\dagger h
-\frac{\bar\lambda}{2}(h^\dagger h)^2.
\label{eff}
\end{eqnarray}
The renormalized Yukawa coupling $\bar g_{t}$, Higgs mass $m_h$ and quartic coupling $\bar \lambda$ are $d=4$ relevant operators in the IR scaling domain, where the composite SM Higgs boson (wave-function renormalized) behaves as an elementary SM Higgs boson. The covariant derivative $D_\mu$ is the same as that 
in the SM. The $\Delta L_{\rm gauge}$ and $L_{\rm kinetic}$ are the usual SM Lagrangians of gauge bosons and fermions. All renormalizable ($d=4$ relevant) operators receive 
fermion-loop contributions, 
define for the low-energy scale $\mu$ and follow RG scaling laws. The $\Delta L_{\rm irr}$ represents the 
non-renormalizable ($d>4$ irrelevant) operators induced by four-fermion operators (\ref{bhllow}) with SM interactions and propagators.

The conventional renormalization $Z_\psi=1$ for fundamental 
fermions and the unconventional wave-function renormalization (form factor)
$\tilde Z_h$ for the composite SM Higgs 
boson $h$ are adopted
\begin{equation}
\tilde Z_{h}(\mu)=\frac{1}{\bar g^2_t(\mu)},
%\, \bar g_t(\mu)=\frac{Z_{HY}}{Z_H^{1/2}}g_{t0}; 
\quad \tilde \lambda(\mu)=\frac{\bar\lambda(\mu)}{\bar g^4_t(\mu)}.
%\,\bar\lambda(\mu)=\frac{Z_{4H}}{Z_H^2}\lambda_0,
\label{boun0}
\end{equation}
%where $Z_{HY}$ and $Z_{4H}$ are proper renormalization constants of the Yukawa coupling and quartic coupling in Eq.~(\ref{eff}). 
In the IR scaling domain, the full one-loop RG
equations for running couplings $\bar g_t(\mu^2)$ and $\bar \lambda(\mu^2)$ read %\cite{Bardeen1990}
\begin{eqnarray}
16\pi^2\frac{d\bar g_t}{dt} &=&\left(\frac{9}{2}\bar g_t^2-8 \bar g^2_3 - \frac{9}{4}\bar g^2_2 -\frac{17}{12}\bar g^2_1 \right)\bar g_t,
\label{reg1}\\
16\pi^2\frac{d\bar \lambda}{dt} &=&12\left[\bar\lambda^2+(\bar g_t^2-A)\bar\lambda + B -\bar g^4_t \right],\quad t=\ln\mu \label{reg2}
\end{eqnarray}
where one can find $A$, $B$ and RG equations for 
SM $SU_c(3)\times SU_L(2)\times U_Y(1)$ running  renormalised gauge couplings $\bar g^2_{1,2,3}(\mu^2)$ in Eqs.~(4.7), (4.8) of 
Ref.~\cite{Bardeen1990}. 
The SSB-generated top-quark mass $m_t(\mu)=\bar g_t^2(\mu)v/\sqrt{2}$. 
One describes the 
$\langle\bar tt\rangle$-composite Higgs-boson by its pole mass $m^2_h(\mu)=2\bar \lambda(\mu) v^2$, form factor $\tilde Z_h(\mu)=1/\bar g_t^2(\mu)$ and effective quartic coupling $\tilde\lambda(\mu)$, provided that
$\tilde Z_h(\mu)>0$ and $\tilde\lambda(\mu)>0$ are obeyed.

\begin{figure*}[t]
%\begin{center}
\hskip0.9cm\includegraphics[height=3.40in]{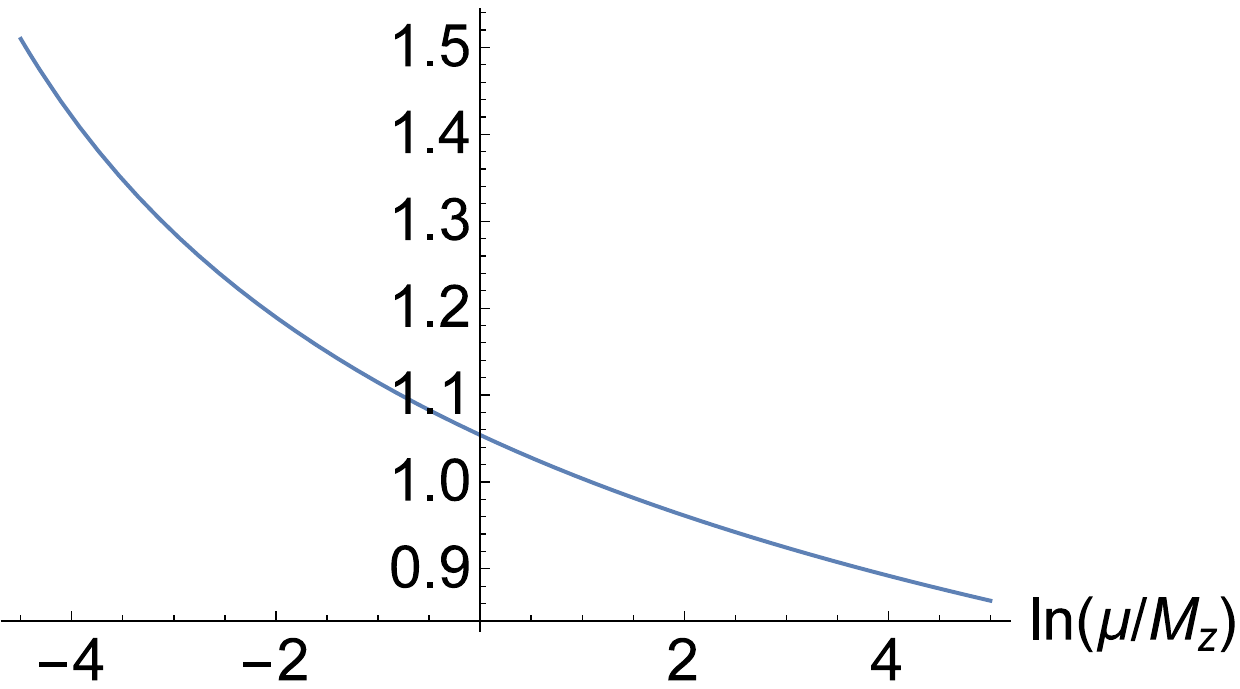}\hskip1.6cm\includegraphics[height=3.40in]{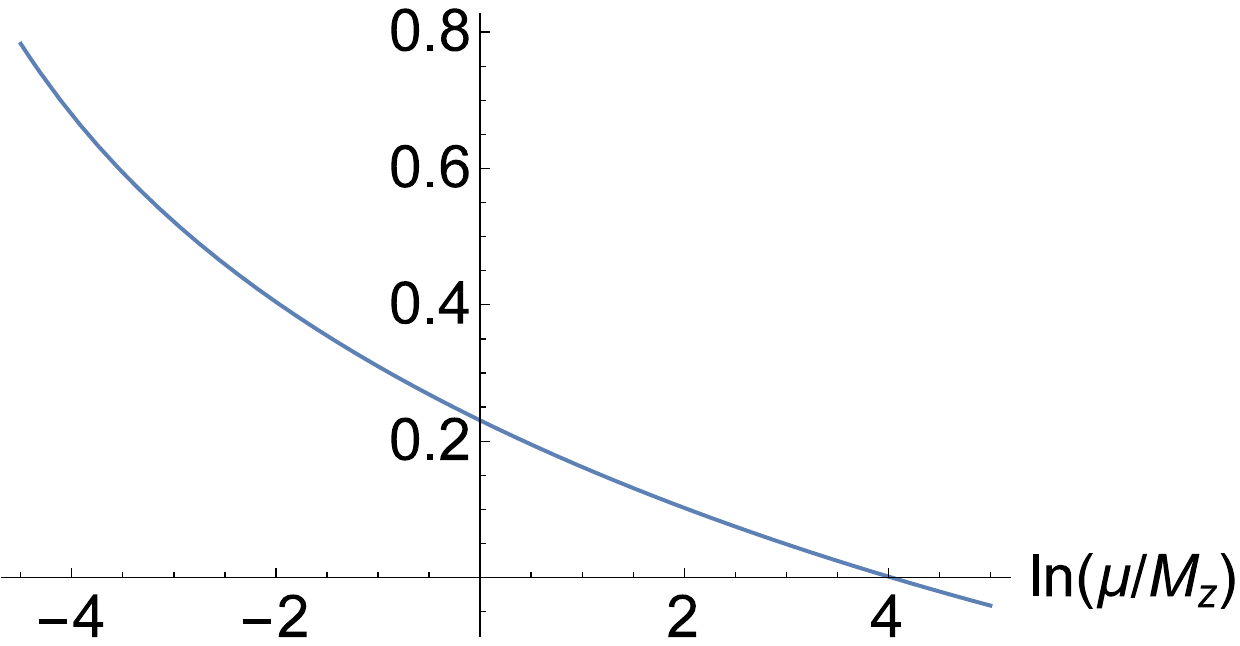}
\put(-115,175){$\tilde\lambda(\mu)$}
\put(-340,175){$\bar g_t(\mu)$}
\vskip-1.1in
\caption{In the top-quark channel, after renormalization, the effective Higgs Yukawa coupling $\bar g_t(\mu)$ and quartic coupling 
$\tilde\lambda(\mu)$ as functions of energy scale $\mu$ are determined by RG equations (\ref{reg1},\ref{reg2}), 
mass shell condition  (\ref{thshell}) of the experimentally measured top quark and Higgs mass. Figure left panel shows the top-quark Yukawa coupling is about unity at the SM scale $v$ and slowly decreases to $0.9$ at TeV scales. The Yukawa coupling represents a finite form factor $\tilde Z_h(\mu)=1/\bar g_t^2(\mu)$ of composite SM Higgs boson. The 
$\tilde Z_h(\mu)$ increases with energy, indicating the composite SM Higgs boson becomes a tighter and tighter bound state as energy increases. 
After wave-function renormalization, the composite SM Higgs boson behaves as an elementary SM Higgs boson. Figure right panel shows the effective Higgs quartic coupling $\tilde\lambda(\mu)$, which relates to the pole mass $m_h$ (\ref{thshell}) of the composite Higgs boson. It shows the energy-running Higgs mass decreases with the energy scale increases. The $\tilde\lambda(\mu)$ value is about $0.15$ at the SM scale $v$, and decreases as energy increases and becomes negative at the energy scale $\approx 5.1$ TeV.
We reproduce these figures from Refs.~\cite{Xue2013, Xue2014}.}
\label{figy}
%\end{center}
\end{figure*}

To definitely solve the RG equations (\ref{reg1}) and (\ref{reg2}) for 
$\bar g_t$ and $\bar\lambda$, so as to obtain masses $m_t$ and $m_h$, it requires boundary conditions at a physical energy scale. BHL adopt the operator (\ref{bhl}) and the theoretical compositeness conditions,
\begin{eqnarray}
\tilde Z_H(\Lambda_{\rm cut})=1/\bar g_t^2(\Lambda_{\rm cut})=0;\quad
\tilde\lambda(\Lambda_{\rm cut})=0,
\label{bhlc}
\end{eqnarray}
at the cutoff scale $\Lambda_{\rm cut}$, where the effective Lagrangian (\ref{eff}) is sewed together with the underlying four-fermion Lagrangian (\ref{bhl}). However, such 
solutions $\bar g_t(\mu)$ and $\bar\lambda(\mu)$  
cannot reproduce %simultaneously 
correct experimental values of the masses $m_t$ and $m_{h}$. It implies that the conditions (\ref{bhlc}) should be incorrect.  

Instead, we consider the operator (\ref{bhllow}) and the IR fixed point $G_c=8\pi^2/(N_c\Lambda^2)$ at the scale $\Lambda$ and use the mass-shell conditions of 
experimental values $v$, $m_t$ and $m_h$
\begin{eqnarray}
m_t(m_t)=\bar g_t^2(m_t)v/\sqrt{2}\approx 173 {\rm GeV},
\quad m_h(m_h)=[2\bar \lambda(m_h)]^{1/2} v\approx 126 {\rm GeV},
\label{thshell}
\end{eqnarray}
as the boundary conditions for the RG Eqs.~(\ref{reg1}) and (\ref{reg2}).
As a result we find  \cite{Xue2013,Xue2014} the solutions $\bar g_t(\mu)$
and $\tilde\lambda(\mu)$ (Fig.~\ref{figy}) that are different from the solutions 
(Figs.~4 and 5) in Ref.~\cite{Bardeen1990}. 
In the IR domain of low energies $\mu\gtrsim M_z$, for $d=4$ renormalizable operators, 
the effective Lagrangian (\ref{eff}) and 
RG equations (\ref{eff}-\ref{reg2}) %(\ref{eff},\ref{boun0},\ref{reg1},\ref{reg2}) 
with experimental boundary conditions (\ref{thshell}) are effectively equivalent 
to the SM Lagrangian 
and RG equations of elementary top-quark and Higgs field. However, we have to mention in the effective Lagrangian (\ref{eff}) there are 
non-renormalizable ($d>4$ irrelevant) operators $\Delta L_{\rm irr}$ upon the ground state of spontaneous symmetry breaking. These operators do not follow RG scaling laws and are suppressed by powers of $v/\Lambda$. They are $(d>4)$ 
1PI vertex functions induced by the four-fermion interactions (\ref{bhllow}) 
together with other SM interactions and particle propagators. These operators $\Delta L_{\rm irr}$ in the IR domain are beyond the renormalizable SM Lagrangian. They can give corrections in powers of $v/\Lambda$ to the SM results. Because the scale $\Lambda$ is only about one order of magnitude larger than the electroweak scale $v$, 
their leading order contributions can be possibly sizable and significant for precision measurements in experiments.

\subsection{Effective theory of composite particles at TeV scale}

Because we use the boundary condition (\ref{thshell}), the $\bar g_t(\mu)$ and $\bar\lambda(\mu)$ values at the SM scale $v$ agree with the SM phenomenological values. However, the energy dependence of effective Yukawa coupling $\bar g_t(\mu)$
and quartic coupling $\bar\lambda(\mu)$ in Fig.~\ref{figy} show (i) how they vary as functions of low energy $\mu$ within the IR scaling domain for the SM; (ii) what their behaviours are when energy increases and goes beyond the SM. In the latter case, 
the validity of the effective low-energy SM Lagrangian (\ref{eff}) and the perturbative one-loop RG equations (\ref{reg1}) and (\ref{reg2}) is in question. Therefore, we regard the $\bar g_t(\mu)$ and $\bar\lambda(\mu)$ for $\mu\gg v$ in Fig.~\ref{figy} as high-energy extrapolations from their low-energy solutions for $\mu\sim {\mathcal O}(v)$. Observe that the $\bar\lambda(\mu)$ becomes negative around $5.1$ TeV, and the ground-state energy of 
effective SM Lagrangian (\ref{eff}) has no bound 
from below. Namely, the SM Hamiltonian ${\mathcal H}\propto - L \propto \frac{\bar\lambda}{2}(h^\dagger h)^2$ is negative and is not bounded from below, and the SM is unstable against Higgs field quantum fluctuations. In an alternative view, Higgs boson becomes a tachyon for its mass $m_h=(2\bar \lambda)^{1/2}v$ runs into imaginary.
This inconsistency indicates the breakdown of
the low-energy SM Lagrangian (\ref{eff}) and RG scaling laws (\ref{reg1}) and (\ref{reg2}) in the IR domain. It implies the previously discussed physics of composite particles' theory in the UV domain beyond the SM in the IR scaling domain occurs about the TeV scale. The massive composite bosons $H$ (\ref{doublet}) have masses $M_H\propto \Lambda$ (\ref{chm}), and proportionality is about the order of unity.

\begin{figure*}[t]
\centering
\includegraphics[height=6.0cm,width=13.8cm]{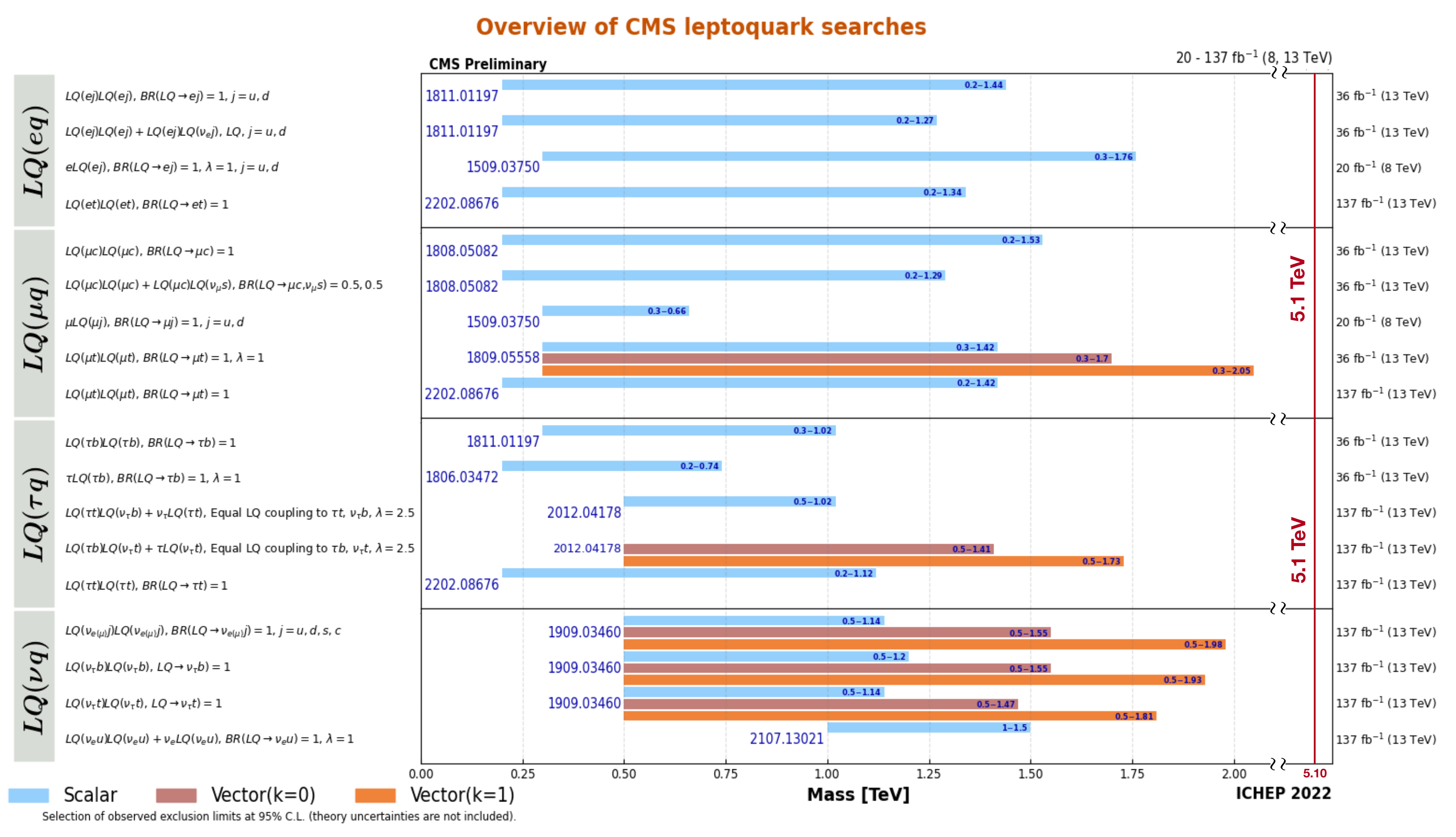}
%\includegraphics[height=3.40in]{xue-cms5.10-deep-red}
%\vskip-1.1in
\caption{We reproduce this table from the Leptoquark summary plot of 
ICHEP 2022 meeting (https://twiki.cern.ch/twiki/bin/view/CMSPublic/SummaryPlotsEXO13TeV). 
It summarises the experimental leptoquark searches in masses (TeV) and channels. We insert the red vertical line to indicate the typical scale $\Lambda\approx 5.1$ TeV of the effective theory of composite particles, e.g., the massive composite Higgs bosons $H$ and Leptoquark scalars, whose masses are characterized by $\Lambda$.}
\label{figscale}
\end{figure*}

The scale $5.1$ TeV is an approximate value in the order of magnitude since it achieves by extrapolating low-energy solution $\bar\lambda(\mu)$ to the high-energy regime.  
We compare this value with the existing bound from current experiments.
There is no any experimental study of massive composite bosons $H$.
We find many experimental studies of the Lepto-quark (LQ) boson states see Fig.~\ref{figscale}. 
%https://twiki.cern.ch/twiki/bin/view/CMSPublic/SummaryPlotsEXO13TeV#Leptoquark_summary_plot 
In this scenario, the scalar LQ states form in the same dynamics and UV domain as the massive composite bosons $H$ when we use the lepton-quark four-fermion interactions, see Sec.~IV A 3 of Ref.~\cite{Xue2017}. Namely, the massive composite LQ bosons form via the lepton-quark four-fermion interactions, while the massive composite $H$ bosons form via the quark-quark four-fermion interactions (\ref{rs2}). 
The massive composite LQ bosons and massive composite $H$ bosons have the same characteristic mass scale $M\propto\Lambda$.
The current experimental studies on the massive composite LQ bosons, see the blue (scalar) in Fig.~\ref{figscale} give some insight into the lower bounds of their mass scales (TeV). These lower bounds also apply to the masses $M_H$ of massive composite bosons $H$ and the characteristic scale $\Lambda$ of the effective theory of composite particles 
in the UV domain. 
Figure \ref{figscale} shows the value $\Lambda\approx 5.1$ TeV (red vertical line) has not yet 
been reached. In connection with the LHC experiment searches, we currently perform phenomenological investigations of massive composite LQ bosons and fermions in this scenario.

\section{Extrapolation from IR domain to UV domain}\label{extrap}

Using the experimental results (\ref{thshell}) and RG solutions (Fig.~\ref{figy}) 
of the SM field theory (\ref{eff}) in the IR scaling domain at the scale $v$, we attempt to qualitatively infer the scale $\Lambda$ for the effective field theory (\ref{effH}) in the UV scaling domain.
The latter associates strong four-fermion coupling dynamics and requires non-perturbative approaches. The same question is for the QCD dynamics too. What is the connection between perturbative QCD and non-perturbative QCD? The former is in the UV scaling domain for asymptotically free quarks and gluons at the scale $\Lambda_{\rm QCD}$. The latter is in the IR scaling domain for hadronic bound states at hadron mass scales \cite{Deur2016}. 

The RG equations (\ref{reg1}) and (\ref{reg2}) are obtained by perturbative calculations using the smallness of four-fermion couplings in the IR scaling 
domain. Therefore the solutions (Fig.~\ref{figy}) should be quantitatively valid only for low energies. Nevertheless, to gain an insight into their qualitative behaviour in high energy, 
we extrapolate the solutions (Fig.~\ref{figy}) to the high energy range  $\mu\gg M_z$. 
As a result, we find that (i) the Yukawa coupling $\bar g_t(\mu)$ is the order of unity, the form factor $\tilde Z_h(\mu)\not=0$ is finite, and the $\langle\bar tt\rangle$-composite Higgs boson is a tightly bound state, behaves 
as if an elementary Higgs boson after the 
wave-function renormalisation;
(ii) the effective quartic coupling $\tilde\lambda(\mu)$ decreases as the scale $\mu$ increases and becomes negative at the energy scale of $5.1$ TeV. It indicates when the energy scale $\mu$ is more than $5.1$ TeV, the SSB dynamics and ground state break down since the total energy does not bound from below. It implies the new dynamics and ground state of strong-coupling four-fermion interactions. Therefore, it infers the characteristic 
scale $\Lambda\approx 5.1$ TeV for the gauge symmetric field theory of composite particles beyond the SM.

The non-perturbative Wilson renormalisation group (WRG) needs to be used to describe the connection between the IR and UV 
scaling domains. The WRG flow from the theory (\ref{eff}) for the SM in the IR scaling domain runs towards 
the theory (\ref{effH}) for composite particles in the UV 
scaling domain. The boundary conditions for the WRG flow should associate with the symmetry-breaking phase transition and energy-gap ground state change at the scale $\Lambda$.
We expect that for the $\langle\bar tt\rangle$ 
channel, the SM effective Lagrangian (\ref{bhllow}) of $\langle\bar tt\rangle$-composite 
Higgs boson $h$ 
is sewed together with the effective Lagrangian (\ref{effH}) of the composite boson $H_t^0$ 
at the scale $\Lambda$. It achieves
by connecting the Higgs boson $h$ 
form factor $\tilde Z_h$ to the composite boson $H_t^0$ form factor $Z_H$, namely $\tilde Z_h\rightarrow Z_H$.
%, thus Yukawa couplings $\bar g_t\sim g_H$ should be in the same order. 

Since the scale $\Lambda$ is not very much larger than the scale $v$, and ${\mathcal O}(v^2/\Lambda^2)\sim 10^{-2}$, no extreme fine-tuning needs to achieve the electroweak scale $v$ from the scale $\Lambda$ 
of the effective theory of composite particles.
Moreover, in the low-energy effective Lagrangian (\ref{bhllow}), we introduce the $d>4$ irrelevant operators $\Delta L_{\rm irr}$ suppressed by at least $(v/\Lambda)$ in the IR domain. One of them is the effective right-handed couplings of $W^\pm$ and $Z^0$, e.g., $\Gamma_\mu^W= i\frac{\bar g_{2}}{\sqrt{2}}\gamma_\mu P_R\,{\mathcal G}^W_R$ and ${\mathcal G}^W_R\sim (v/\Lambda)$, see operators (5.1,5.6,5.7) of the previous article \cite{Xue2022b}. These operators are absent in the SM Lagrangian. They possibly give sizable corrections to SM observable, for example, the $W$ and $Z$ gauge bosons' masses and decay widths. In Ref.~\cite{Xue2022b}, we calculated their corrections to the $W$ boson mass $M_W$ that can account for the recent high-precision measurement of the $M_W$ in $7.0~\sigma$ disagreement with the Standard Model expectation \cite{Aaltonen2022}, and we also calculated their corrections to $W$and $Z$ gauge bosons' decay widths for future experimental verification.
 
\section{Phenomenological implications}\label{phenoW}

\subsection{Composite SM Higgs boson in IR domain at electroweak scale}

First, we discuss the possible phenomenological difference of composite SM Higgs in this scenario from elementary SM Higgs in the low-energy IR domain. Their composite and elementary natures are not distinguishable around the electroweak scale
$v$. They have the same kinematics, 
SM gauge couplings, Yukawa and quartic couplings. The only differences are  
the energy-scale dependence of effective top-quark Yukawa coupling $\bar g_t(\mu)$ and quartic couplings $\bar\lambda(\mu)$ given 
in Fig.~\ref{figy}. We expect Yukawa and quartic couplings' values to depart from the SM values when the energy scale $\mu$ significantly increases, leading to possible observable effects.

Due to its $\bar tt$-composite nature, the composite SM Higgs boson couples $\bar g_t(q)$ to a top-quark loop. The composite SM Higgs boson has the same production and decay processes as the elementary SM Higgs boson. Its production cross-section and decay width are proportional to $\bar g^2_t(q)$, where $q=\mu$ is the Higgs boson energy. 
They are the same as the SM counterparts around $\bar g^2_t(m_t)\approx 1.04$ at the top-quark mass $m_t$. The Higgs boson production cross-section and decay rate decrease when the Higgs boson energy $q$ increases. 
However, Fig.~\ref{figy} shows  
$\bar g^2_t(m_t)\approx 1.04$ and   
%$\bar g^2_t(2m_t)\approx 0.98$,$\bar g^2_t(3m_t)\approx 0.95$, 
$\bar g^2_t(4m_t)\approx 0.93$ varies 
slowly. We expect the deviation of Higgs decay width from the SM one to be 
$11\%$ for $\Delta \bar g^2_t\approx 0.11$ or the decay-rate ratio $\Gamma_H/\Gamma_H^{\rm SM}\approx \bar g^2_t(4m_t)/\bar g^2_t(m_t)\approx 0.89$. We presented similar discussions in Sec.~III A 2 of Ref.~\cite{Xue2017}.
Such discrepancy is still too small to be clearly 
distinguished by the present precision level of LHC experiments, $\Gamma_H/\Gamma_H^{\rm SM}\approx 1.11^{+0.63}_{-0.60}$ and $\Gamma_H\approx 4.6^{+2.6}_{-2.5}$ MeV, see Michiel Veen's presentation on behalf of the ATLAS Collaboration in Higgs 2022 Pisa. 

The Higgs quartic coupling $\lambda(\mu)$ as a function of the energy scale is fundamental for fully understanding the physics of the SM and beyond. Figure \ref{figy} (right) shows the Higgs quartic coupling $\lambda(\mu)$ more quickly decreases as the energy scale increases, compared with the Yukawa coupling $\bar g_t(\mu)$. However, the $\lambda(\mu)$-measurement via triple Higgs final states is out of reach even for HL-LHC, and none of the present or future colliders is known to be able to determine the $\lambda$ coupling \cite{Plehn2005}. Nevertheless, relating to the running Higgs mass $m_h(\mu)=[2\bar\lambda(\mu)]^{1/2}v$ (\ref{thshell}), one may have some insight into the $\lambda(\mu)$ coupling as a function of energy scale by measuring the triple Higgs coupling $\lambda(\mu) v$, see N.~De Filippis' presentation in 
Higgs 2022 Pisa provided enough experimental precision achieves.

In this scenario, spontaneous SM gauge symmetry breaking yields 
only top-quark mass $m_t$ or Yukawa coupling $\bar g_t$ \cite{Bardeen1990}, and such configuration of the symmetry-breaking ground state is energetically favourable \cite{Preparata1996, Xue2013c}. Other SM fermions masses or Yukawa couplings ($m_f=\bar g_f v$)
are produced by explicit SM gauge symmetry breaking via the three quarks and leptons family mixing with the top quark \cite{Xue:2016dpl}. The smallness of other fermions' masses is because of the smallness of their mixing with the top quark. Namely, their Yukawa couplings are much smaller than $\bar g_t\sim {\mathcal O}(1)$. SM neutrinos $\nu_L^\ell$ ($\ell=e,\mu,\tau$) and sterile right-handed neutrinos $\nu^\ell_R$ (a kind of dark matter particle) also 
receive small Dirac masses and Yukawa couplings to the composite SM Higgs boson. The branching ratio of Higgs decay via these invisible channels (missing energy momenta) should be small. 
In addition, the composite SM Higgs boson also 
has 1PI vertices 
coupling to other kinds of dark matter particles 
$\chi$boson and axion $a$, which are scalar and 
pseudo-scalar composite states of sterile neutrinos \cite{Xue2022a}. These are high dimensional ($d>4$) operators in $\Delta L_{\rm irr}$ of the effective Lagrangian (\ref{eff}). We expect their contributions to Higgs invisible decay channels to be small and suppressed by powers of $v/\Lambda$. It needs some detailed investigations.

\subsection{Massive composite bosons in UV domain at TeV scale}

To end this article, 
we discuss some peculiar properties and gauge couplings of massive composite bosons $H$. 
They are distinct from the $\langle\bar tt\rangle$-composite Higgs for the SM. It provides phenomenological implications for experimental searches.
In the UV scaling domain, via the Yukawa $g_H$ couplings in the effective Lagrangian (\ref{effH}), massive composite bosons $H$ decay to a left-handed quark and a right-handed quark, and the decay rate is $\Gamma_H \approx (3/16\pi)g_H^2M_H$. 
They can decay to two SM gauge bosons, for example, via a triangle diagram of a massless quark loop \cite{Xue2017},
\begin{eqnarray}
H^0_{t,b}\rightarrow \gamma\gamma;\quad Z^0Z^0;\quad \gamma Z^0;\quad W^+W^-
\label{Hdecay}
\end{eqnarray}
and two gluons, as well as charged channel
$H^\pm\rightarrow W^\pm +\gamma (Z^0)$.

In the UV domain, like as photon $\gamma$ ($A_\mu$), the electroweak gauge bosons $Z^0$ and $W^{\pm}$ are 
massless due to exact $SU_L(2)\times U_Y(1)$ gauge symmetries. For these massless gauge fields in the UV scaling domain, we adopt the SM gauge fields' combinations $W^{\pm}\equiv (W^1\mp i W^2)/\sqrt{2}$,
\begin{eqnarray}
Z^0\equiv -B\sin\theta_\chi +W^3\cos\theta_\chi,\quad A \equiv B\cos\theta_\chi+ W^3 \sin\theta_\chi,
\label{wzmassl}
\end{eqnarray}
where $W^{1,2}$ $(W^3)$ is the charged (neutral) 
$SU_L(2)$ field and $B$ is the $U_Y(1)$ field. The angle $\theta_\chi$ %($\tan \theta_\chi=g_1/g_2$) 
mixes neutral fields $(W^3,B)$ to represent physical fields $(\gamma,Z^0)$, and charge $e=g_2\sin\theta_\chi=g_1\cos\theta_\chi$. The massless gauge bosons on the mass-shell $q^2=0$ differ from massive gauge bosons on the mass-shell $q^2=M^2_{W,Z}$. The former propagators mediate a long-range electroweak force between composite particles. While the latter propagators mediate a short-range electroweak force between SM elementary particles. 
For the sake of simplicity, we use in Eq. (\ref{wzmassl}) the notations for massless $W^\pm$ and $Z^0$, which are the same 
as massive counterparts in the SM. 

Moreover, the massless neutral gauge fields (\ref{wzmassl}) can differ from their massive counterparts in the SM since the  $\cos^2\theta_\chi$ could be different from the Weinberg $\cos^2\theta_W$ value. Recall that one obtains Weinberg angle $\cos^2\theta_W= (M_W/M_Z)^2$ by diagonalising the gauge-boson mass matrix in the IR scaling domain. However it does not depend on the electroweak scale $v$, and it depends only on $SU_L(2)$ and $U_Y(1)$
gauge couplings $\cos^2\theta_W= g_2^2/(g_1^2+g_2^2)$. The SM $SU_c(3)\times SU_L(2)\times U_Y(1)$
gauge group couplings $g_3$, $g_2$ and $g_1$ are the same in the IR and UV domains. The electric charge $e$ is a definite number for both elementary and composite particle spectra. However, the $g_2$ and $g_1$ values can change because the spectrum (group representation) changes from elementary particles to composite particles in the phase transition between the UV and IR domains. 
If this is the case, 
$\cos^2\theta_\chi\not=\cos^2\theta_W$. We expect the difference should be small, of the order of $(v/\Lambda)^2$, and the weak first-order transition should be smoothly continuous. 
The $\cos^2\theta_\chi$ value should be fixed by experiments for massive composite particles and massless gauge bosons in the UV scaling domain 
at the TeV scale.

In the UV scaling domain for massive composite particles, 
the SM chiral gauge symmetries are not broken spontaneously, therefore the SM elementary fermions are massless 
and there are no Goldstone modes. 
The massless $W^\pm$ and $Z^0$ bosons (\ref{wzmassl}) at high energies (TeV) behave exactly like the photon and have two transverse components and no longitudinal one. Thus they do not decay to two SM fermions, in contrast to the SM massive $W^\pm$ and $Z^0$ decay channels, e.g.~$W^\pm\rightarrow \ell^\pm + \nu_\ell$ of the rate $\Gamma_W= (3/16\pi) g_2^2 M_W$. Therefore, the final state and kinematics of massive composite bosons (resonances) $H$ decay to $Z^0Z^0$ and $W^+W^-$ (\ref{Hdecay}) are not 
four-lepton states, e.g., $\ell\ell\nu\nu$. They should be back-to-back events like the two-photon $\gamma\gamma$ channel. These massless 
$W^\pm$ and $Z^0$ have interacting vertexes with the left- and 
right-handed components of SM elementary fermions according to their quantum numbers of SM gauge symmetries. Therefore, their detection should be in the same way as detecting photons.

Analogously to the Berit-Wheeler process $\gamma+\gamma\rightarrow \ell^+\ell^-$, two massless gauge bosons $W^\pm$, $Z^0$ and $\gamma$ collisions produce a lepton pair \footnote{ 
They can produce two massive composite particles, provided their energies are larger than two composite particle masses.}. Thus the massless $W^\pm$ and $Z^0$ from $H^0_{t,b}$ and $H^\pm$ decay (\ref{Hdecay}) could be detected by the channels $W^\pm + \gamma^*\rightarrow \ell^\pm + \nu_\ell (\bar\nu_\ell)$ and $Z^0 + \gamma^*\rightarrow \ell^+ +\ell^-$,
in analogy to $\gamma + \gamma^*\rightarrow \ell^+ +\ell^-$.  The $\gamma^*$ represents virtual photons around a nucleus inside the target. The 
cross-section should be proportional to the Berit-Wheeler cross-section $\sigma_{BW}\approx \pi (\alpha/{\mathcal E})^2$ at the limit of high energy ${\mathcal E}\gg m_\ell$, where $m_\ell$ is the 
lepton mass. For example, $\sigma_{W\gamma}\approx \sigma_{BW}/(2\sin^2\theta_\chi)$ for the channels $W^\pm + \gamma^*\rightarrow \ell^\pm + \nu_\ell (\bar\nu_\ell)$, final states are left-handed leptons, and the energy ${\mathcal E}$ is at the TeV scale. 
In addition, the resonances $H$ and their decays (\ref{Hdecay}) should play roles in
vector-boson-fusion (VBF) processes at high energies of the TeV scale. There is no unitarity problem in vector-boson-scattering (VBS) since massless $W^\pm$ and $Z^0$ have no longitudinal components. In other words, the longitudinal components of SM massive $W^\pm$ and $Z^0$ gauge bosons decouple when the symmetry-restoring phase transition occurs at the TeV scale. The present scenario of massive composite particles and massless gauge bosons at the TeV scale modifies the thermal history of electroweak symmetry breaking in cosmology. Detailed investigations will be the subjects of future studies. 

In phenomenology, it is necessary to compare and contrast the particles ($H,h$) with two Higgs-type resonances of mass terms $m^2$ and $M^2\gg m^2$ on the SSB ground state $\langle\phi\rangle\not=0$, like two-Higgs-doublet models \cite{Branco2012} and the model discussed in Ref.~\cite{Consoli2022}.
One of the distinct features is that the composite bosons $H$ (Table \ref{qQF0}) are doublets of SM gauge-symmetry quantum numbers, 
interacting accordingly with the massless SM gauge bosons. The interactions and spectra of massive composite particles and massless gauge bosons fully preserve the SM chiral gauge symmetries in the UV scaling domain.

\comment{
Since the $\langle\bar tt\rangle$ composite Higgs $h$ (or equivalently elementary Higgs Yukawa coupling to $\langle\bar tt\rangle$) and composite boson $H_t^0$ have the same constituents $\langle\bar tt\rangle$, SM quantum numbers and gauge couplings. Therefore, they should have the same interacting and decaying channels. This implies two resonances of Higgs type at different mass scales $M_H,m_h$, and decay rates $\Gamma_H\sim M_H,\Gamma_h\sim m_h$.  % due to $M_H > m_h$ giving a large phase space.  
}

\vskip0.1cm
\section{Acknowledgment.}  
The author thanks Sehar Ajmal, 
Jethro Gaglione, Alfredo Gurrola,
Matteo Presilla, Orlando Panella, Francesco Romeo and Hao Sun for many discussions. 
The author also thanks the Editor and Referee for their reviewing the manuscript.

%\bibliography{../Axion-like}
%\bibliography{Axion-like}

\providecommand{\href}[2]{#2}\begingroup\raggedright\endgroup

\end{document}